\documentclass[12pt]{article}
\usepackage{amsmath}
\usepackage{psfrag,epsf,graphicx}
\usepackage{enumerate}
\usepackage{natbib}
\usepackage{url} 
\usepackage{microtype}

\usepackage{svg}

\newcommand{\blind}{0}

\usepackage{multirow,subcaption,amsfonts,stmaryrd,amssymb,amsthm,mathtools,bm,tikz,pifont,XCharter,booktabs,tikz-cd}
\usepackage[dvipsnames]{xcolor}
\usepackage[ruled]{algorithm2e}
\usetikzlibrary{calc,bayesnet}
\usepackage[backref=page]{hyperref}
\hypersetup{
    colorlinks=true,
    citecolor=blue,
}
\newcommand*{\cond}{\!\mid\!}
\newcommand*{\dquote}[1]{``{#1}"}
\newcommand*{\tr}{\mathsf{T}}

\newcommand*{\ind}{\stackrel{ind}{\sim}}
\newcommand*{\iid}{\stackrel{iid}{\sim}}
\newcommand*{\set}[1]{\left\{{#1}\right\}}
\newcommand*{\bracket}[1]{\left({#1}\right)}
\newcommand*{\sqbracket}[1]{\left[{#1}\right]}

\theoremstyle{definition}

\newtheorem{define}{Definition}[section]
\newtheorem{example}{Example}[section]

\usepackage{listings}
\definecolor{codegreen}{rgb}{0,0.6,0}
\definecolor{codegray}{rgb}{0.5,0.5,0.5}
\definecolor{codepurple}{rgb}{0.58,0,0.82}
\definecolor{backcolour}{rgb}{0.95,0.95,0.92}
\lstdefinestyle{mystyle}{
    backgroundcolor=\color{backcolour},   
    commentstyle=\color{codegreen},
    keywordstyle=\color{magenta},
    numberstyle=\tiny\color{codegray},
    stringstyle=\color{codepurple},
    basicstyle=\ttfamily\footnotesize,
    breakatwhitespace=false,         
    breaklines=true,                 
    captionpos=b,                    
    keepspaces=true,                 
    numbers=left,                    
    numbersep=5pt,                  
    showspaces=false,                
    showstringspaces=false,
    showtabs=false,                  
    tabsize=2
}
\begin{document}

\def\spacingset#1{\renewcommand{\baselinestretch}%
{#1}\small\normalsize} \spacingset{1}


\if0\blind
{
  \title{\bf
    Efficient scenario analysis in \\ real-time Bayesian election forecasting \\ via sequential meta-posterior sampling
}
  \author{Geonhee Han\hspace{.2cm} \\
    \small Department of Statistics, Columbia University\thanks{gh2610@columbia.edu.} \\
    \small Department of Statistics, University of Tokyo\thanks{GH was supported by the JST SPRING GX fellowship program.}
    \\ \\
    Andrew Gelman \hspace{.2cm} \\
    \small Department of Statistics and Department of Political Science, Columbia University
    \\ \\
    Aki Vehtari \hspace{.2cm} \\
    \small Department of Computer Science, Aalto University
    }
\date{4 Aug 2026}
  \maketitle
\bigskip
\begin{abstract}

Bayesian aggregation lets election forecasters combine diverse sources of information, such as state polls and economic and political indicators:
as in our collaboration with {\em The Economist} magazine.
However, the demands of real-time posterior updating, model checking, and communication introduce practical methodological challenges.
In particular, sensitivity and scenario analysis help trace intricate dependencies and understand model behavior.
Yet, under standard Markov chain Monte Carlo, even small tweaks to the model (e.g., in priors, data, hyperparameters) require full refitting, making it computationally expensive in real time.
We introduce a meta-modeling strategy paired with a sequential sampling scheme;
by traversing posterior meta-models, we enable real-time structured analyses without repeated refitting.
In a back‑test of the model, we demonstrate substantial computational gains, and show {how seemingly innocuous data‑wrangling choices can unintentionally introduce partisan asymmetries into the model's mechanisms for handling systematic polling error}.
Code is available at \href{https://github.com/geonhee619/SMC-Sense}{https://github.com/geonhee619/SMC-Sense}.

\end{abstract}
} \fi

\if1\blind
{
  \bigskip
  \bigskip
  \bigskip
  \begin{center}
    {\LARGE\bf Efficient scenario analysis in \\ real-time Bayesian election forecasting \\ via sequential meta-posterior sampling}
\end{center}
  \medskip
\bigskip
\begin{abstract}

Bayesian aggregation lets election forecasters combine diverse sources of information, such as state polls and economic and political indicators:
as in our collaboration with {\em The Economist} magazine.
However, the demands of real-time posterior updating, model checking, and communication introduce practical methodological challenges.
In particular, sensitivity and scenario analysis help trace intricate dependencies and understand model behavior.
Yet, under standard Markov chain Monte Carlo, even small tweaks to the model (e.g., in priors, data, hyperparameters) require full refitting, making computationally expensive in real time.
We introduce a meta-modeling strategy paired with a sequential sampling scheme;
by traversing posterior meta-models, we enable real-time structured analyses without repeated refitting.
In a back‑test of the model, we demonstrate substantial computational gains, and show {how seemingly innocuous data‑wrangling choices can unintentionally introduce partisan asymmetries into the model's mechanisms for handling systematic polling error}.
Code is available at \href{https://github.com/geonhee619/SMC-Sense}{https://github.com/geonhee619/SMC-Sense}.

\end{abstract}
} \fi

\noindent%
\vfill

\newpage


\spacingset{1.1}
\renewcommand{\arraystretch}{0.6}

\newpage

\section{Introduction} \label{sec:1}
\paragraph{Bayesian aggregation for election forecasting.}

Bayesian aggregation provides a method to combine various sources of information for election forecasting.
These include the large number of publicly available pre-election polls  at the state- and national-level \citep{HolbrookDeSart1999, Linzer2013}, economic indicators \citep{Campbell1992, GelmanKing1993, WlezienErikson1996, BartelsZaller2001, NadeauLewis-Beck2001, EriksonWlezien2008}, and
historical voting patterns \citep[e.g.,][]{LockGelman2010}.
In this paper, we use the forecasting model developed with the {\em Economist} magazine \citep{morris2020, HeidemannsGelmanMorris2020, GelmanGoodrichHan2024}, an extension of the hierarchical Bayesian framework first introduced by \cite{Linzer2013}, designed to infer the day-by-day evolution of state-level opinions
while simultaneously accounting for sampling error, nonsampling polling error, and variation between states.

\paragraph{The methodological challenge.}

This work aims to address a methodological challenge that arises at the intersection of large‑scale data fusion, daily updating, model checking, and communication.
The model is estimated in real-time via \texttt{Stan} \citep{CarpenterGelmanHoffmanLeeGoodrichBetancourtBrubakerGuoLiRiddell2017}, and refines its forecasts daily as new opinion polls become available.
That the model is never \textit{final}, where each new data point prompts revisions to both the forecasts and the model itself \citep{GelmanGoodrichHan2024}, necessitates methods for routine model checks which help justify revisions in ways that are flexible, scalable, and communicable.

This is the case in all statistical workflows \citep[e.g.,][]{cook1977, cook1979, belsley1980, huber1981, rosenbaum1983, eubank1984, carnegie2016}.
They are especially salient here;
we are concerned not only with sensitivity analysis, in which one examines how robust the final inferences are to alternative configurations of model specifications or hyperparameters, \citep[e.g.,][]{Canavos1975, ross1987, Gustafson1996, Clarke1998, BergerInsuaRuggeri2000, ZhuIbrahimTang2011, RoosMartinsHeldRue2015, MasoeroStephensonBroderick2018},
but also with scenario analysis, in which one explores how the final forecasts change under different hypothetical conditions \citep[Chap.~8]{BayesianWorkflow2026}.

\begin{figure}
    \centering
    \subfloat[Baseline posterior, 90 days out]{\includegraphics[width=.49\linewidth]{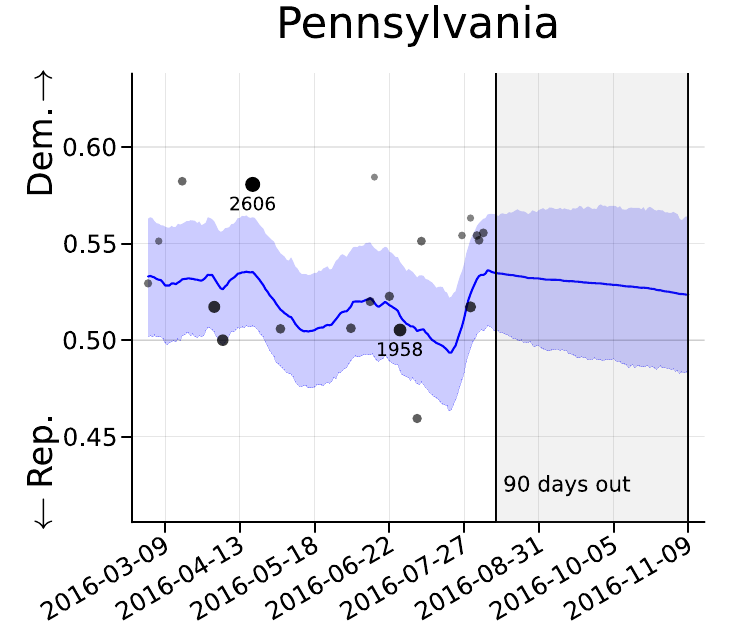}} 
    \subfloat[After hyperparameter perturbation]{\includegraphics[width=.49\linewidth]{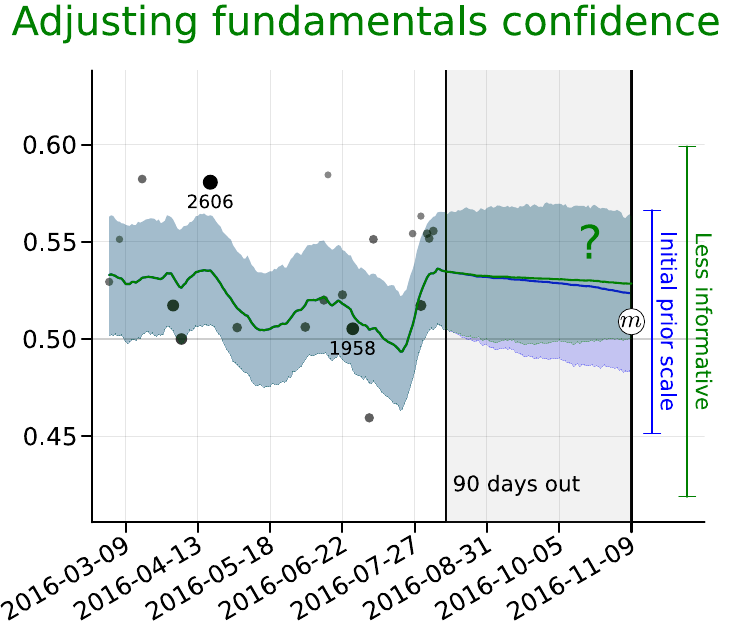}}
    \\
    \spacingset{1.0} \caption{\em
        Panels (a) and (b) illustrate the role of scenario/sensitivity analysis in the real-time forecasting setup.
        (a) exemplifies the baseline posterior before any perturbation.
        (b) demonstrates how modifying hyperparameters may affect inferences and forecasts.
        This process is embedded in a live forecasting system;
        to support routine model checking and forecast revision, we seek computationally efficient methods for timely, time-specific scenario exploration and sensitivity analysis without repeated model refitting.
    } \label{fig:schematic}
\end{figure}

It is computationally demanding to perform such analyses \textit{a posteriori} and in real-time, particularly when relying on conventional Markov chain Monte Carlo (MCMC) methods.
Each incoming batch of data necessitates full posterior recomputation;
any tweaks to quantities the model takes as fixed inputs require model refitting.
Consequently, additional diagnostics by perturbing each data or individual hyperparameter becomes impractical.

The relevant existing computational strategies are summarized in Table \ref{tab:literature}.
Beyond brute force MCMC, deletion-based perturbative methods based on importance reweighting have been proposed for case-influence and out-of-sample predictive assessment \citep{GelfandDey1994, Peruggia1997, EpifaniMacEachernPeruggia2008, VehtariGelmanGabry2017, HanGelman2026}.
These offer approximate case-deleted posterior diagnostics, but are limited in scope for our purposes.
The \textit{infinitesimal} approach provides local sensitivity analysis with respect to hyperparameters or likelihood contributions to the posterior mean \citep{Giordano2018, KallioinenPaanenBurknerVehtari2023, NguyenGiordanoMeagerBroderick2024}.
Although fast and mathematically elegant, these can be difficult to interpret, communicate in collaborative settings, or qualitatively guide specific model refinements.

\begin{table}
    \centering
    \spacingset{1.0}
    \begin{tabular}{p{2.2cm} p{4.4cm} p{3.6cm} p{4.6cm}}
        \textbf{} & \textbf{Use / pros} & \textbf{Mismatch} & \textbf{Literature (e.g.)} \\
        \toprule
        Brute force MCMC & Generic, interpretable,
        
        easy to implement & Inefficient
        
        re-computation \\
        \midrule
        Deletion reweighting & Case deletion
        
        (e.g., case influence;
        
        cross-validation
        
        for out-of-sample eval.) & Limited to
        
        case-influence & \cite{GelfandDey1994},
        
        \cite{EpifaniMacEachernPeruggia2008},
        
        \cite{VehtariGelmanGabry2017}.
        \\
        \midrule
        Infinitesimal & Local and linear sensitivity to hyperparameter or prior/likelihood contribution & Can be hard to
        
        interpret and
        
        communicate & \cite{Giordano2018},
        
        \cite{KallioinenPaanenBurknerVehtari2023},
        
        \cite{NguyenGiordanoMeagerBroderick2024}. \\
        \midrule
        Sequential & Hyperparameter sensitivity; cross-validation & Want broader perturbation schemes
        
        (e.g., data arrival) & \cite{BornnDoucetGottaro2010},
        
        \cite{HanGelman2026}. \\
        \midrule
        \begin{tabular}[t]{@{}l@{}}
            \textbf{This work}
        \end{tabular}
         & \multicolumn{3}{l}{
            \begin{tabular}[t]{@{}l@{}}
                Sequentially re-cycle computation for efficiency. \\
                Applicable/applied to real-time scenario/sensitivity analysis. \\
                Interpretable; ease of communication. \\
                Modular; no code rewriting per scheme \\ (\texttt{BridgeStan} accesses \texttt{Stan} internals).
            \end{tabular}
        }
        \\
    \end{tabular}
    \spacingset{1.0} \caption{\em
        Summary of potentially applicable existing computational approaches.
        The listed literature is illustrative and not exhaustive.
        \textbf{Abbreviation}:
            MCMC = Markov chain Monte Carlo.
    } \label{tab:literature}
\end{table}

\paragraph{Scope.}

Our goal is to develop a strategy for post-MCMC scenario and sensitivity analysis, compatible with the real-time sequential forecasting context.
In particular, sequential Monte Carlo (SMC) samplers \citep{DelMoralDoucetJasra2006} provide a foundation for interpretable posterior diagnostics through user-controlled hyperparameter variation \citep[e.g.][]{BornnDoucetGottaro2010}.
We adapt this approach to further address the specific demands of our setting, including
sequential real and hypothetical data arrivals,
time-specific sensitivity diagnostics, and
time-specific what-if exploratory analysis.

\paragraph{Structure.}

Section \ref{sec:2} presents the data description, formal model specification, and the specific diagnostic scenario/sensitivity queries we seek to address.
Section \ref{sec:3} introduces the strategy and details the implementation, by formulating a \textit{meta-model} paired with a sequential sampling scheme to traverse the meta-model.
Section \ref{sec:4} backtests the model using data from the 2016 U.S.~presidential election.
Section \ref{sec:5} concludes.


\section{Bayesian aggregation for election forecasting} \label{sec:2}
\subsection{{Data: pre-election polls, fundamentals, and state demographics}} \label{sec:model:data}

{
The data used in the forecasting model combine
(a) state and national pre‑election polls which cover the full campaign period from around March through Election Day,
(b) historical national economic and political indicators, and
(c) state‑level demographic and geographic information.
The data used specifically in this work are sourced from \href{https://github.com/TheEconomist/us-potus-model}{this repository}, for the year 2016.
The choice is discussed in detail in Section \ref{sec:model:q}, where we explain how the year provides the conditions relevant for the model stress tests.
}

{
The polling data in item (a) include aggregate information such as
the target (state or national),
pollster,
sample size,
population type (e.g., likely voters, adults),
survey mode (e.g., online, live phone),
field dates, and
the preprocessed two‑party vote shares for the Democratic and Republican candidates.
A binary indicator $\iota_i$ is also coded for \dquote{adjuster} pollsters known to weight their samples to match population demographics,
for the model to apply partisan non‑response adjustment.
}

{
Indicators in item (b), namely, Q2 GDP growth, June presidential approval, and past presidential election results, serve as a prior on the national- and state-level vote shares, via the incumbent party's expected national vote share based on the \dquote{Time for Change} model \citep{Abramowitz1988}.
At the state-level, historical partisan deviation of each state from the national average is computed using the results of the previous presidential election. 
}

{
The demographic and geographic data in (c) include
historical Democratic vote share,
state demographic distributions (e.g., population, density, racial composition, education, age),
urbanicity, and
the share of white evangelical voters.
These are used to construct an empirical correlation matrix across states, later used by the model to encode inter-state dependence.
Here, negative correlations are truncated at zero to favor national uniformity, and the empirical matrix is composed with an equi-correlation matrix using a mixing parameter $\lambda$ that blends demographic information and a uniform national swing. 
}

\subsection{The model} \label{sec:2:1}

We give a brief overview of how the data are combined in \textit{The Economist} model described in the following for the years 2020 and 2024.
{The core dimensions and indices that map the polling data into the hierarchical structure is given in Table \ref{tab:index}.
A tabulated summary of the variables and notation is given in Table \ref{tab:notation}.}

\begin{table}
    \centering
    \spacingset{1.0}
    \begin{tabular}{p{4.0cm} p{10cm}}
        Index/dimension & Description \\
        \toprule
        $t = 1, \ldots, T$ & Days in the forecasting timeline (through Election Day) \\
        \midrule
        $i = 1, \ldots, N_t, \ldots, N$ & Total number of polls across the full campaign.
        
        $N_t$: national and state‑level polls available up to day $t$. \\
        \midrule
        $s = 1, \ldots, S$ & States \\
        \midrule
        $h[i] \in \set{1, \ldots, H}$ & Polling organization for poll $i$ \\
        \midrule
        $m[i] \in \set{1, \ldots, M}$ & Polling mode for poll $i$ (e.g., live phone, online) \\
        \midrule
        $p[i] \in \set{1, \ldots, P}$ & Population type for poll $i$ (e.g., likely/registered voters) \\
        \midrule
        $t[i] \in \set{1, \ldots, T}$ & Observation day for poll $i$ \\
        \bottomrule
    \end{tabular}
    \spacingset{1.0} \caption{\em
        Dimensions and indices used throughout.
        See Table \ref{tab:notation} in for notation.
    } \label{tab:index}
\end{table}

We now define the likelihood.
Let $y_i$ and $n_i - y_i$ denote the number of respondents in poll $i$ supporting the Democratic and Republican candidates, respectively.  
Both $y_i$ and $n_i$ are observed in the model with
\begin{equation*}
    y_i \ind \text{binomial}(n_i, \pi_i).
    \tag{$i = 1, \ldots, N_t$}
    \label{model:eq:likelihood}
\end{equation*}
The likelihood does not explicitly account for design effects or nonsampling error;
these are partly accounted for by additional error terms in the model \citep{GelmanGoodrichHan2024}.

\begin{table}[!h]
    \centering
    \spacingset{1.0}
    \begin{tabular}{p{2cm} p{3cm} p{9cm}}
        Variable & Index & Description \\
        \toprule
        $y_i$
        
        ($n_i - y_i$) & $i = 1,\ldots,N_t$ & Respondents in poll $i$ supporting the Democratic (Republican) candidate \\
        \midrule
        $\mu_{t,s}$ & $t = 1,\ldots,T$;
        
        $s = 1,\ldots,S$ & Latent time-varying opinion in state $s$ on day $t$ \\
        \midrule
        $e_s$ & $s = 1,\ldots,S$ & Election-level polling bias in state $s$ \\
        \midrule
        $w_s$ & $s = 1,\ldots,S$ & Fixed weight for state $s$ (used to aggregate state trends into national-level predictions) \\
        \midrule
        $\alpha_i$ & $i = 1,\ldots,N_t$ & Nonsampling error for poll $i$ \\
        \midrule
        $\alpha_h^{(h)}$ & $h = 1,\ldots,H$ & House effects \\
        \midrule
        $\alpha_m^{(m)}$ & $m = 1,\ldots,M$ & Mode effects \\
        \midrule
        $\alpha_p^{(p)}$ & $p = 1,\ldots,P$ & Population effects \\
        \midrule
        $\alpha_t^{(b)}$ & $t = 1,\ldots,T$ & Election-level time-varying bias \\
        \midrule
        $\varepsilon_i$ & $i = 1,\ldots,N$ & Idiosyncratic noise \\
        \bottomrule
    \end{tabular}
    \spacingset{1.0} \caption{\em
        Notation used throughout.
        See also Table \ref{tab:index} for dimensions and indices.
    } \label{tab:notation}
\end{table}

{First, each individual poll is assigned indices mapping it to its
observation day $t[i]$,
state $s[i]$,
pollster $h[i]$,
mode $m[i]$,
population type $p[i]$, and
a binary indicator $\iota_i$ for whether the pollster corrects for partisan non‑response.
These enter the latent two‑party support rate for poll $i$ as binomial success probability $\pi_i$.
Here, $\pi_i$ is modeled on the logit scale through an additive structure, based on
the latent support trend in state $s$ at time $t$ ($\mu_{t,s}$),
state-level error term ($e_s$),
poll-specific bias ($\alpha_i$), and
poll-specific error ($\varepsilon_i$),
}
\begin{equation*}
    \operatorname{logit}(\pi_i) =
    \begin{rcases}
    \begin{dcases}
        \mu_{t[i],s[i]} + e_{s[i]} & (\text{\( i \) is state-level}) \\
        \sum_{s=1}^S w_s (\mu_{t[i],s} + e_s) & (\text{\( i \) is national})
    \end{dcases}
    \end{rcases}
    + \alpha_i + \sigma_i\varepsilon_i.
    \tag{$i = 1, \ldots, N_t$}
\end{equation*}
{The fixed weights $(w_s)_s$ represent each state's projected share of the national popular vote based on prior‑cycle turnout (scaled by adult‑population growth).}

Latent trends $\bm{\mu}_t = (\mu_{t,1}, \dots, \mu_{t,S})$ are modeled as a correlated random walk,
\begin{align*}
    \bm{\mu}_{t}
        \sim
        \text{MVN}(\bm{\mu}_{t+1}, \bm{\Sigma}^{(\mu)}).
    \tag{$t = T-1, \ldots, 1$}
\end{align*}
The random walk is specified backward in time so that the fundamentals-based prior anchors the Election‑Day distribution, which incorporates information from economic and political indicators,
\begin{equation*}
    \bm{\mu}_T
        \sim
        \text{MVN}(\bm{m}, \bm{\Sigma}),
\end{equation*}
{where $\bm{m}$ is the fundamentals-based prediction on the logit scale.}

The bracketed term also includes the state-by-state polling errors $(e_s)_s$, which are modeled jointly to capture inter-state correlated biases,
\begin{equation*}
    (e_1, \dots, e_S) \sim \text{MVN}(\bm{0}, \bm{\Sigma}^{(\text{bias})}).
\end{equation*}

{The covariances $(\bm{\Sigma}^{(\mu)}, \bm{\Sigma}, \bm{\Sigma}^{(\text{bias})})$ encode both potential inter-state dependencies and prior uncertainty, and are derived from both empirical correlations such as in the state demographic distributions.
See Section \ref{sec:model:data} and \ref{sec:model:q:corr} for how they are specifically constructed.}

{The poll-specific bias term $\alpha_i$ is decomposed into a sum of components related to known sources of nonsampling variation \citep[e.g.,][]{McDermottFrankovic2003, WlezienErikson2006}:
polling house effects $\alpha_{1:H}^{(h)} \iid \text{normal}(0, v^{(h)})$,
response mode effects $\alpha_{1:M}^{(m)} \iid \text{normal}(0, v^{(m)})$,
polling population effects $\alpha_{1:P}^{(p)} \iid \text{normal}(0, v^{(p)})$,
and an election-level time-specific error $\alpha_{1:T}^{(b)}$ (as a stationary normal mean-zero autoregression of order 1),}
\begin{equation}
    \alpha_i = \alpha_{h[i]}^{(h)} + \alpha_{m[i]}^{(m)} + \alpha_{p[i]}^{(p)} + \iota_i  \alpha_{t[i]}^{(b)}.
    \label{model:eq:errors}
\end{equation}
The indicator $\iota_i \in \set{0,1}$ is active ($\iota_i = 1$) when the pollster $h[i]$ is classified as an adjuster and is inactive otherwise for non-adjusters.
In place of $\iota_i \alpha_{t[i]}^{(b)}$, the 2024 version of the model further adds an adjustment for partisan sponsorship;
given $\eta \sim \text{exponential}(\nu)$ with rate $\nu>0$, the additional term $\alpha_{l[i]}^{(l)}$ takes $+\eta$ for Democratic-sponsored polls, $-\eta$ for Republican-sponsored polls, and 0 otherwise \citep{GelmanGoodrichHan2024}.

Finally, independent measurement errors are modeled as
$\varepsilon_i \iid \text{normal}(0, 1)$.
The scale $\sigma_i = \sigma^{(\text{state})} > 0$
if $i$ is a state poll and
$\sigma_i = \sigma^{(\text{national})} > 0$
if $i$ is a national poll.

Both state and national forecasts evolve dynamically throughout the campaign, in the sense that each day there is an increase $t \gets t+1$ and the number of data points $N_t$ increases as the polling accumulates.
Consequently, the generative model in Section \ref{sec:2:1} defines the joint distribution, specifically on day $t$,
\begin{align}
    p(\bm{\Theta}, y_{1:N_t})
        &=
        {
            \underbrace{\sqbracket{\prod_{h=1}^H p(\alpha_h^{(h)})}}_{\text{House}}
            \underbrace{\sqbracket{\prod_{m=1}^M p(\alpha_m^{(m)})}}_{\text{Mode}}
            \underbrace{\sqbracket{\prod_{p=1}^P p(\alpha_p^{(p)})}}_{\text{Population}}
            \underbrace{p(\alpha_{1:T}^{(b)})}_{\text{Election-level bias}}
        } \,
        \underbrace{p(e_1, \ldots, e_S \cond \bm{\Sigma}^{(\text{bias})})}_{\substack{\text{State-by-state bias}}} \, \nonumber
        \\ &\qquad
        \underbrace{p(\bm{\mu}_T \cond \bm{m}, \bm{\Sigma})}_{\substack{\text{Fundamentals-augmented} \\ \text{Election Day outcome}}}
        \underbrace{\sqbracket{\prod_{t=1}^{T-1} p(\bm{\mu}_{t} \cond \bm{\mu}_{t+1}, \bm{\Sigma}^{(\mu)})}}_{\text{State-by-state trends}}
        \underbrace{\sqbracket{\prod_{i=1}^N p(\varepsilon_i)}}_{\text{Idiosyncratic noise}}
        \underbrace{\sqbracket{\prod_{i=1}^{N_t} \mathbb{P}(y_i \cond n_i, \pi_i, \sigma_i)}}_{\text{Sampling variability}}
        .
        \label{eq:model:joint}
\end{align}
{
We collect the parameters of interest shared throughout the campaign as
\begin{equation}
    \bm{\Theta}
    :=
    (\alpha_{1:H}^{(h)}, \;
    \alpha_{1:M}^{(m)}, \;
    \alpha_{1:P}^{(p)}, \;
    \alpha_{1:T}^{(b)}, \;
    e_{1:S}, \;
    \bm{\mu}_{1:T}, \;
    \varepsilon_{1:N}).
    \label{eq:params}
\end{equation}
Only the first $N_t$ poll-specific errors $\varepsilon_i$ enter the likelihood;
the remaining terms are prior-only.
Errors for $i > N_t$ are still retained in the formulation, as the workflow operates over a consistent parameter space throughout.
In particular, its dimension, namely $N$ (that of the noise $\varepsilon_i$), is independent of $t$ with a fixed $N > N_t$ at all times.
All $\varepsilon_i$ for $i > N_t$ are simply drawn from their prior.
(Fixing $N$ up front is sufficient for the retrospective exercise considered in this work.
In a real-time setting where $N_t \geq N$ may occur, one could always enlarge $N$ by drawing additional $\varepsilon_i$ from the prior as needed.)
}

\subsection{{Scrutinizing the workflow}} \label{sec:model:q}

The rationale for incorporating structure beyond random variability, particularly in the intercepts $\alpha_i$ and their constituents $(\alpha_{1:H}^{(h)}, \, \alpha_{1:M}^{(m)}, \, \alpha_{1:P}^{(p)}, \, \alpha_{1:T}^{(b)})$, is that the generative process of polls is not simple random draws, but contain substantial nonsampling errors that do not cancel out when averaged/aggregated.
Therefore, in producing time‑ and state‑specific latent trends, the intent is to eliminate these systemic biases through Bayesian marginalization:
$\bm{\mu}_{1:T} = (\bm{\mu}_1, \ldots, \bm{\mu}_T)$ for state‑level forecasts
and weighted aggregation, $(\sum_{s=1}^S w_s \mu_{t,s})_{t=1}^T$, for national forecasts.

At the same time, the model and the forecasting workflow also present how the Bayesian aggregator must reconcile the noisy and fragmented information at once, and cues how some part of the model and workflow may be mis-elicited.
Rather than simply treating this as a flaw, we acknowledge it as an expected condition to combine limited and heterogeneous evidence under a high‑dimensional structure.
In the following, we try to actively and specifically re-evaluate the modeling workflow.

\subsubsection*{Vulnerability to election-level polling bias} \label{sec:model:q:bias}

\begin{itemize}
    \item[Q1.] How does the hierarchical model handle internally simulated industry‑wide polling failures, and in what ways is it vulnerable?
\end{itemize}

Pre-election polls are subject not only to sampling errors but also to various nonsampling errors, such as those identified by the \dquote{total survey error} framework: including frame, nonresponse, measurement, and specification errors \citep{Biemer2010}.
International evidence from electoral surveys indicates that the observed mean absolute errors are larger than can be attributed to sampling variability alone \citep{JenningsWlezien2018}.
Empirical evaluations of historical polling data in the U.S.~demonstrate systemic election-level bias that can be shared between polls within a given election \citep{Shirani-MehrRothschildGoelGelman2018}.
This is in part due to the shared methodological challenges faced by polling organizations;
for example, individual pollsters may face similar difficulties in reaching specific demographic subgroups \citep[differential nonresponse:][]{Gelman2021}.
Since these are correlated across different pollsters, they often cannot be eliminated merely by aggregating multiple surveys, and importantly shape the extent to which the forecaster can rely on polls in relation to other available sources of information.

A widely discussed example is the 2016 cycle, where public surveys understated support for one candidate in several key swing states \citep{GelmanAzari2017}.
Subsequent analyses have attributed part of the discrepancy to differential nonresponse \citep{KennedyBlumenthalClementClintonDurandFranklinMcGeeneyMiringoffOlsonRiversSaadWittWlezien2018, GelmanHullmanWlezienElliot2020}.

In this context, recall that the hierarchical model assigns a prior expectation of zero to uniform polling biases $(e_s)_s$.
Because the magnitude of such cycle-wide errors would not be identifiable from polling data alone in a given cycle, there is a risk that the model may effectively be blind to them.
This motivates our first perturbation exercise: to verify this vulnerability and examine how the hierarchical structure absorbs (or fails to absorb) these deviations.

\subsubsection*{Covariance and prior-data conflict} \label{sec:model:q:corr}

\begin{itemize}
    \item[Q2.] How and in what ways are the model's posterior geographic inferences sensitive to the forecaster's weighting about spatial covariance?
\end{itemize}

\begin{figure}
    \centering
    \includegraphics[width=.8\linewidth]{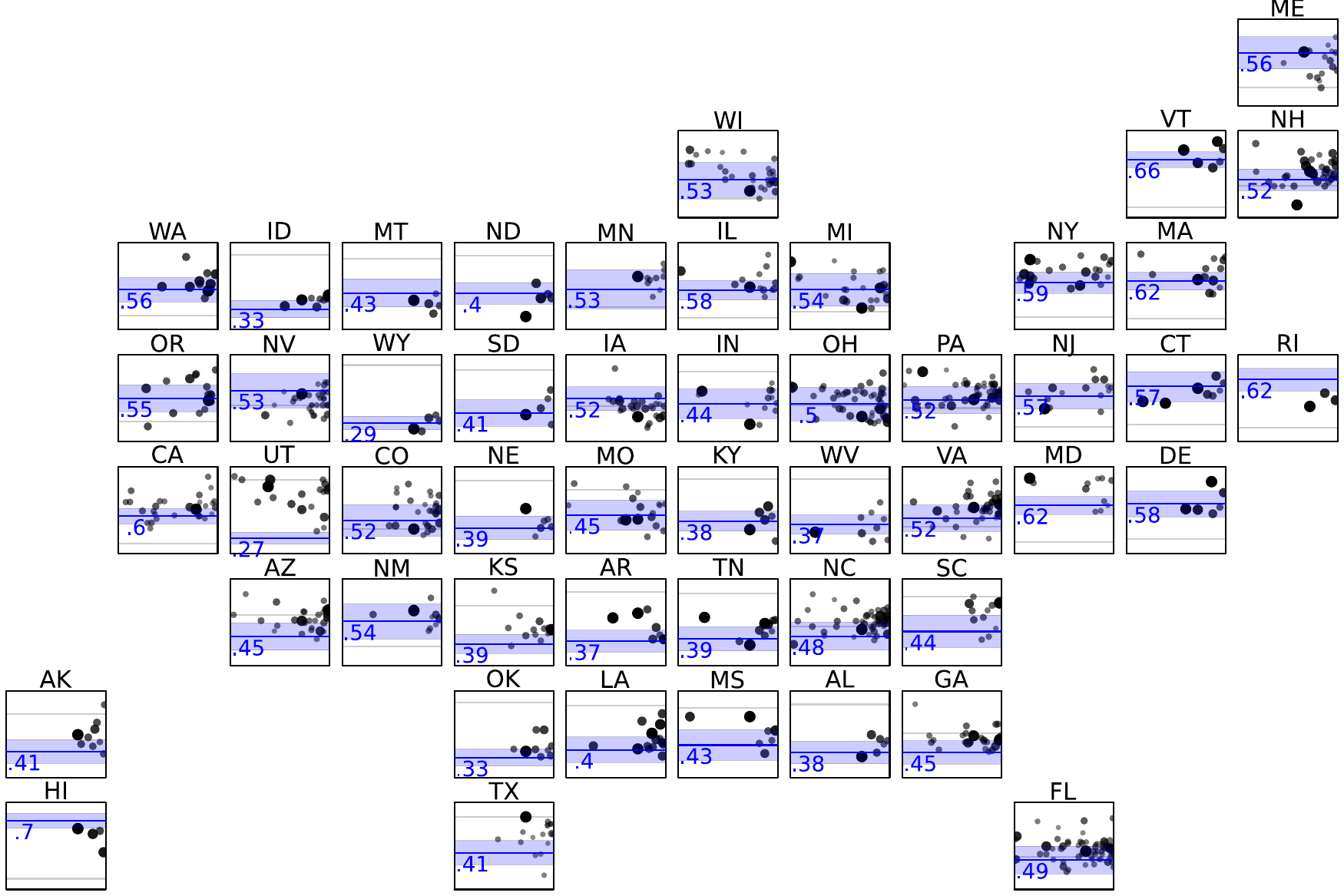}
    \spacingset{1.0} 
    \caption{\em
        State-level panels showing polling from March to 14 days before the 2016 election (\textbf{black dots}, proportional to sample size $n_i$).
        Fundamentals‑based predictions (and 2-SD prior band) are shown as the \textbf{\textcolor{blue}{blue line}} (and the \textbf{\textcolor{blue}{shaded region}}).
        The gray horizontal line marks the 0.5 threshold.
    }
    \label{fig:2016}
\end{figure}

Under the presence of persistent polling bias, another source of concern is from the model's treatment of prior-data conflict, particularly via the fixed relationship between the fundamentals-based prior and the polling data.
Also in 2016, the fundamentals prior implied a rather tight race, whereas the polling data showed that the Democratic candidate performed substantially better: see Figure \ref{fig:2016}.

The persistent inequality between heterogeneous inputs creates a prior-data conflict that the model must reconcile throughout the campaign.
Latent trends are centered on the fundamentals \textit{a priori} but can move toward polling data through the state-level covariance structure (Section \ref{sec:2:1}).
This, in the workflow, is specified as a mixture of a demography-based correlation $\bm{R}$ and an equi-correlation matrix $\bm{R}_1$, as $\lambda \bm{R} + (1-\lambda) \bm{R}_1$, with a forecaster‑chosen heuristic weight $\lambda = 3/4$.
This weight controls how easily the model can shift all states upward in tandem to accommodate the polling-prior gap.

This way in which the workflow specifies state-by-state covariances has not been systematically validated.
The data-processing decision in constructing $\bm{R}$ and therefore $\bm{\Sigma}^{(\mu)}$ and $\bm{\Sigma}$ after marginal scaling (Section \ref{sec:model:data}), although apparently minor, is a subjective modeling choice in the covariance structure with potentially large downstream effects, because the resulting geographic reconciliation of the prior-data conflict depends directly on how this mixture is specified, given the form $(\bm{\mu}_T - \bm{m})^\tr \bm{\Sigma}^{-1} (\bm{\mu}_T - \bm{m})$.
The second perturbation therefore investigates the sensitivity, to clarify to what extent the forecasts are driven by incidental tuning.

\subsubsection*{Asymmetries in data collection} \label{sec:model:q:asymmetry}

\begin{itemize}
    \item[Q3.] How differently does the high-dimensional hierarchical model absorb and propagate new localized polling shocks across states under varying geographic and metadata configurations?
\end{itemize}

The geographic and temporal distribution of the polling data is shaped by economic and institutional incentives rather than random sampling \citep{GelmanHullmanWlezienElliot2020}.
Campaigns operate in a constrained environment that creates incentives in how they select between different activities \citep{GoldsteinPaula2002, WangLewisSchweidel2018, GordonLovettLuoReeder2023};
for example, there is little strategic incentive to target safe states over \dquote{battlegrounds} \citep{ShacharNalebuff1999, GimpelKaufmannPearson-Merkowitz2007}.
Similarly, survey organizations, also facing substantial costs, are incentivized to concentrate their efforts in competitive states, leaving others sparsely sampled.
Furthermore, public surveys can be designed primarily for media consumption rather than for the precise measurement needs of researchers \citep{BarberMannMonsonPatterson2014}.
As a result of many such incentives, new information is systemically prone to (apparent) shocks, and they enter the model asymmetrically over the electoral map, even when they can largely be artifacts of noise.

Regardless, the model is designed such that these enter as fresh data points that must be reconciled with the existing posterior, via hierarchical partial pooling and the spatial covariance structure that shall smooth estimates across states.
Here, it remains unclear whether the model maintains the intended behavior in response to localized deviations.
On the one hand, the spatial prior or the idiosyncratic error term may naturally resist allowing a central state to move independently, so its many unshocked neighbors act as a regularizer that suppresses localized surprises.
On the other hand, the model's weighting on uniform national swings may make it more probable under the prior to undesirably shift the entire map slightly rather than isolate the shocked state, in which case shocks may further propagate outward.
In the third perturbation exercise, our aim is to examine how the model absorbs, propagates, and explains polling surprises under different geographic contexts and metadata conditions.

\section{Sequential meta-modeling} \label{sec:3}
\subsection{The meta-model} \label{sec:2:2}

To facilitate the scenario and sensitivity analyses framed in Section \ref{sec:model:q} for the above model, we introduce and work directly with a meta-model.
The meta‑model provides a unified parameterization of all perturbations --- prior, likelihood, data, and synthetic insertions --- within a fixed parameter space.
{The idea is presented in general terms in a way that is applicable to a broad class of polling models and thus supports ongoing and future model refinement, rather than committing to a single fixed specification.
Later, we specialize to our specific setting in Sections \ref{sec:3:1} and \ref{sec:4}.}

{
First, let the prior factorize into conditionally independent components $p_{\phi_k}(\bm{\theta}_k)$ over $k = 1, \ldots, K$, and similarly let each data unit (i.e., poll) $i = 1, \ldots, N_t$ contribute a conditionally independent likelihood term $p_{\psi_i}(\bm{y}_i \cond \bm{\Theta})$.
$\phi_k$ denotes the hyperparameters for the $k$-th prior component, and
$\psi_i$ denotes the hyperparameters for the $i$-th likelihood.
The posterior (on day $t$) in terms of the components $\bm{\theta}_{1:K}$ of the full vector $\bm{\Theta}$ is then
\begin{equation*}
    p(\bm{\Theta} \cond \bm{y}_{1:N_t}) \propto
    \sqbracket{\prod_{k=1}^K p_{\phi_k}(\bm{\theta}_k)}
    \sqbracket{\prod_{i=1}^{N_t} p_{\psi_i}(\bm{y}_i \cond \bm{\Theta})}.
\end{equation*}
For our polling model, the prior factors are naturally identified as the individually factorized terms as in Eq.~\ref{eq:model:joint} (e.g., $\bm{\theta}_1 := \alpha_1^{(h)}$, $\bm{\theta}_2 := \alpha_2^{(h)}$, and so on).
}

A further generalization is made to define a meta-model, indexed by an array $\bm{\delta}_t$ encapsulating fixed model quantities, defined as follows.
Let $\rho_k > 0$ be the power coefficient applied to the prior distribution on $\bm{\theta}_k$,
$f_i(\cdot)$ the data perturbation function applied to the $i$-th unit $\bm{y}_i$,
$p_{\psi^*}(\cdot)$ the likelihood for hypothetically inserted data $\bm{y}^*$ (which may consist of one or more observations) with hyperparameter $\psi^*$,
and $\gamma_i, \gamma^* \geq 0$ the power coefficients applied to the likelihood for observed and hypothetical data, respectively.
We define
\begin{align*}
    \bm{\delta}_t
     &:= (\bm{y}_{1:N_t}, \, \phi_{1:K}, \, \rho_{1:K}, \, \psi_{1:N_t}, \, \gamma_{1:N_t}, \, \bm{y}^*, \, \psi^*, \, \gamma^*) \\
    &\mapsto
        \sqbracket{
            \prod_{k=1}^K
            p_{{\phi_k}}(\bm{\theta}_k)^{{\rho_k}}
        }
        \sqbracket{
            \prod_{i=1}^{N_t}
        p_{{\psi_i}}({f_i}(\bm{y}_i) \cond \bm{\Theta})^{{\gamma_i}}
        }
        p_{{\psi^*}}({\bm{y}^*} \cond \bm{\Theta})^{{\gamma^*}}
        .
\end{align*}
The original posterior is then recovered as a special case of the meta-model when
$\rho_k = \gamma_i = 1$ and $\gamma^* = 0$,
with all hyperparameters $(\phi_{1:K}, \psi_{1:N_t})$ held fixed.

Operating over the configuration vector $\bm{\delta}_t$ is therefore one way to retrieve hypothetical posteriors under varying configurations of
hyperparameter settings,
data-value perturbations,
synthetic data insertions, and additionally
prior and likelihood informativity,
while the perturbation path keeps the parameter space fixed.
The idea applies whenever the general factorization holds, and such a decomposition is typically straightforward to obtain for parametric Bayesian hierarchical models.

\subsection{Perturbation path} \label{sec:3:1}

Given the meta-model, and the setup where refreshed initial posterior draws would be affordable/available on a daily basis, the computational problem now is to efficiently traverse a family of posteriors indexed by $\bm{\delta}_t$ about the default level.
We first explicitly define some terminology.

\begin{define}
    Let the {\em baseline} distribution at day $t$ be defined as the posterior under the unperturbed configuration $\bm{\delta}_{t}(0)$: $p(\bm{\Theta} \cond \bm{\delta}_{t}(0))$ (the original one at time $t$).
\end{define} \label{def:baseline}

We then introduce the notion of a perturbation \citep{Weiss1996} as an action that modulates the configuration from $\bm{\delta}_{t}(0)$ to a perturbed setting.

\begin{define}
    A {\em perturbation} $g(\cdot;u)$ is a multiplicative adjustment applied to the unnormalized posterior, summarizing all modulation from $\bm{\delta}_{t}(0)$ to $\bm{\delta}_{t}(u)$, such that
    \begin{equation*}
        p(\bm{\Theta} \cond \bm{\delta}_{t}(u))
        \propto p(\bm{\Theta} \cond \bm{\delta}_{t}(0)) \cdot g(\bm{\Theta}; u),
    \end{equation*}
    producing a continuum of posteriors indexed by $u \in (0,1]$.
\end{define} \label{def:perturbation}

The continuum $\bracket{p(\cdot \cond \bm{\delta}_{t}(u))}_{u \in (0,1]}$ then defines the posterior trajectory under controlled perturbation.
The following are concrete examples to illustrate some modulations relevant to our forecasting setup.

\begin{example}[Sequential updates]
    We want to capture the natural evolution of the posterior as new observations arrive.
    Therefore, in transitioning from $t$ to $t+1$ after receiving new poll data $\bm{y}^* := (y_i)_{i=N_t+1}^{N_t+h}$ with the corresponding metadata $\psi^*$, we first identify the target as
    \begin{equation*}
        p(\bm{\Theta} \cond \bm{\delta}_{t+1}(0)) = p(\bm{\Theta} \cond \bm{\delta}_{t}(1)),
    \end{equation*}
    in which the $N_{t+1}$-th poll is included in $\bm{\delta}_{t+1}(0) = \bm{\delta}_t(1)$.
    This can be bridged from the baseline $p(\bm{\Theta} \cond \bm{\delta}_{t}(0))$ by, for example,
    \begin{equation*}
        g(\bm{\Theta};u)
        := p_{\psi^*}(\bm{y}^* \cond \bm{\Theta})^{\gamma^*(u)}
        \stackrel{\text{e.g.}}{=} \prod_{i=N_t+1}^{N_t+h} \text{binomial}(y_i \cond n_i, \pi_i(\bm{\Theta}))^{\gamma^*(u)},
    \end{equation*}
    where $\gamma^*$ increases from $0$ to $1$ for gradual data injection.
\end{example}

\begin{example}[Hypothetical data insertions]
    This is handled analogously to sequential updates, but with hypothetical data $(\bm{y}^*, \psi^*)$.
    A concrete example is
    \begin{align*}
        \bm{y}^*(u) &= y_{N_t+1}(u), \\
        {y_{N_t+1}(u) \over n_{N_t+1}(u)} &= {1 \over 2}, \\
        \psi^*(u) &\stackrel{\text{e.g.}}{=} \begin{bmatrix}
                \texttt{pollster} \\
                \texttt{mode} \\
                \texttt{population}
            \end{bmatrix}
            = \begin{bmatrix}
                \text{NBC} \\
                \text{Live phone interview} \\
                \text{Registered voters}
            \end{bmatrix}.
    \end{align*}
    (The above is schematic, whereas $\psi^*(u)$ includes more information.)
    Importantly, the rationale here is to allow posterior scenario analysis on counterfactual polls.
\end{example} \label{example:insertion}

\begin{example}[Prior \textit{what-if} \& sensitivity]
    We can define, for a chosen $k$,
    \begin{equation*}
        g(\bm{\Theta}; u)
        :=
        {p_{\phi(u)}(\bm{\theta}_k) \cdot p_{\phi_k}(\bm{\theta}_k)^{-1}}
        ,
    \end{equation*}
    where $\phi(u)$ interpolates between the baseline and alternative prior settings.
\end{example} \label{example:prior}

\begin{example}[Data-value \textit{what-if} \& sensitivity]
    Defining
    \begin{equation*}
        g(\bm{\Theta}; u)
        := {p_{\psi_i}(f(\bm{y}_i;u) \cond \bm{\Theta}) \cdot p_{\psi_i}(\bm{y}_i \cond \bm{\Theta})^{-1}}
        ,
    \end{equation*}
    where $f(\cdot;u)$ is a deterministic transformation of $\bm{y}_i$ that preserves the binomial support and metadata,
    enables local sensitivity analysis to perturbations in the observed poll outcome.
\end{example}

\begin{example}[Data informativity \& prior diffuseness]
    Although not used in our setup, we can modulate the likelihood contribution via $\gamma_i \in (0,1]$
    and similarly the prior via $\rho_k \in (0,1]$
    to control the influence of individual observations and tune the prior strength.
\end{example}

\subsection{Traversing the meta-model: sequential sampling} \label{sec:3:2}

{The above setup is then naturally amenable to a sequential sampling strategy via SMC, without repeated per-configuration simulation.}

As refreshed posterior draws are affordable at least on a daily basis (Section \ref{sec:1}),
initial draws would be available from the baseline distribution in the sense of Definition \ref{def:baseline} on day $t$,
\begin{equation*}
    \bm{\Theta}^{(r)} \sim p(\cdot \cond \bm{\delta}_{t}(0)),
    \tag{$r = 1, \ldots, R$}
\end{equation*}
where $R$ denotes the number of posterior draws.
{(More generally, it is possible to take the prior $p(\cdot \cond \bm{\delta}_{0}(0))$ when $N_{t=0}=0$ and deducing/designing an appropriate sequence of perturbations allow characterizing a continuum of real/hypothetical posteriors over time.)}

Explicitly prepare the mesh,
\begin{equation*}
    0 = u_{\ell=0} < u_{\ell=1} < \ldots < u_{\ell=L} = 1,
\end{equation*}
and exploit the fact that the space $\bm{\Theta}$ is shared along the path; 
{we can exploit the sequential proximity of neighboring configurations to traverse the joint space}
\begin{equation*}
    (\bm{\Theta}_{0}, \bm{\Theta}_{1}, \ldots, \bm{\Theta}_{L})
    \equiv
    \bracket{
        \bm{\Theta} \cond \bm{\delta}_{t}(u_0), \;
        \bm{\Theta} \cond \bm{\delta}_{t}(u_1), \;
        \ldots, \;
        \bm{\Theta} \cond \bm{\delta}_{t}(u_L)
    }.
\end{equation*}

{
Specifically, a sequence of backward Markov kernels \citep[see Eq.~30]{DelMoralDoucetJasra2006} is artificially constructed based on $(\bm{\Theta} \cond \bm{\delta}_{t}(u))$-invariant MCMC kernels for any fixed $u \in (0,1]$ and thus for the full sequential mesh $(u_\ell)_{\ell=1}^L$.
In our case, the ability to obtain high-quality initial MCMC draws from $(\bm{\Theta} \cond \bm{\delta}_{t}(0))$ with dynamic Hamiltonian Monte Carlo \citep{HoffmanGelman2014} implies that the same kernel can be adapted to target all intermediate $(\bm{\Theta} \cond \bm{\delta}_{t}(u_\ell))$ by simply adjusting the configuration to the prescribed $\bm{\delta}_{t}(u_\ell)$.
The unnormalized incremental weights \citep[Eq.~31]{DelMoralDoucetJasra2006} at the $\ell$-th step using only the previous step $(\bm{\Theta}_{\ell-1}^{(r)})_r$ is then expressible in terms of the perturbations,
\begin{equation*}
    \widehat{W}_{\ell}^{(r)}
        = {p(\bm{\Theta}_{\ell-1}^{(r)}, \bm{y}_{1:N_t} \cond \bm{\delta}_{t}(u_{\ell})) \over p(\bm{\Theta}_{\ell-1}^{(r)}, \bm{y}_{1:N_t} \cond \bm{\delta}_{t}(u_{\ell-1}))}
        =
        {g(\bm{\Theta}_{\ell-1}^{(r)}; u_\ell) \over g(\bm{\Theta}_{\ell-1}^{(r)}; u_{\ell-1})}
    .
\end{equation*}
Here, fast-mixing MCMC kernels are helpful in practice but not strictly required; invariance however is essential.
We refer the reader to \cite{DelMoralDoucetJasra2006} 
for a detailed discussion of the construction.}

Similarly to \cite{BornnDoucetGottaro2010} and \cite{HanGelman2026}, the approach differs from bruteforce MCMC in that the MCMC kernel (i.e., rejuvenation) is only invoked when the weights exhibit insufficient sample diversity.
This is measured using the approximate effective sample size of \citet{KongLiuWong1994} and the generalized Pareto shape diagnostic \citep{VehtariSimpsonGelmanYaoGabry2024} (see Section \ref{sec:3:3:diagnostics}).
The result is a sampling-based approximation along the perturbation path that leverages the sequential proximity of adjacent perturbations rather than re‑running full MCMC at each step, {and under which expectations along the perturbed sequence converge to the true target expectation in the limit of infinitely many particles}.
We summarize the procedure in Algorithm \ref{alg:smc}.

\begin{center}
\spacingset{1.1}
\begin{minipage}{0.95\linewidth}
\begin{algorithm}[H]
    \KwIn{MCMC draws $\bm{\Theta}^{(1)}, \dots, \bm{\Theta}^{(R)} \sim p(\cdot \cond \bm{\delta}_{t}(0))$,
    Perturbation schedule $(g(\cdot;u_\ell))_{\ell=1}^L$,
    ESS threshold $\tau \in (0,R)$.
    }
    \medskip
        Initialize particles $(\bm{\Theta}_0^{(r)}, W_0^{(r)}) \gets (\bm{\Theta}^{(r)}, 1/R)$ \;
        \medskip
        \For{$\ell = 1, \ldots, L$}{
            Compute $\widehat{W}_{\ell}^{(r)} \gets g(\bm{\Theta}_{\ell-1}^{(r)}; u_\ell) / g(\bm{\Theta}_{\ell-1}^{(r)}; u_{\ell-1})$ \;
            Compute $W_{\ell}^{(r)} \propto W_{\ell-1}^{(r)} \cdot \widehat{W}_{\ell}^{{(r)}}$ \;
            \uIf{$\texttt{ESS}(W_{\ell}^{(1)}, \ldots, W_{\ell}^{(R)}) < \tau$}{
                $A_{\ell}^{{(r)}} \sim \texttt{Resample}(W_{\ell}^{(1)}, \ldots, W_{\ell}^{(R)})$ \tcp*{Resampling}
                $\bm{\Theta}_{\ell}^{(r)} \gets \texttt{MCMC}(\cdot \cond {\bm{\Theta}}_{\ell-1}^{(A_{\ell}^{{(r)}})})$ \tcp*{$p(\cdot \cond \bm{\delta}_t(u_\ell))$-invariant}
                $W_{\ell}^{(r)} \propto 1$ \;
            }
            \Else{
                $\bm{\Theta}_{\ell}^{(r)} \gets \bm{\Theta}_{\ell-1}^{(r)}$ \;
                \medskip
                $\hat{k}_{\ell} \gets \texttt{ParetoSmooth}(\widehat{W}_{\ell}^{(1)}, \ldots, \widehat{W}_{\ell}^{(R)})$ \tcp*{Diagnostic}
                \uIf{$\hat{k}_{\ell} > 0.7$}{Output warning}
            }
            \medskip
            $\hat{\eta}_{\ell} \gets \sum_{r=1}^R W_{\ell}^{(r)} \eta(\bm{\Theta}_{\ell}^{(r)})$ \;
        }
    {\bf return} Sequential sampling result $(\hat{\eta}_{\ell}, \; (W_{\ell}^{(r)})_r, \; (\bm{\Theta}_{\ell}^{(r)})_r)_{\ell=1}^L$ \;
    
    \caption{\texttt{SMC for scenario/sensitivity analysis}}
    \label{alg:smc}
\end{algorithm}
\end{minipage}
\end{center}


\subsection{Existing approach and potential use cases}

We finally acknowledge some connections to existing approaches.
The perturbation view (in Section \ref{sec:3:1}), and the choice of target function $f$ (in Section \ref{sec:3:2}) provide a unification to re-interpret and extend prior methods.

\paragraph{Data insertion.}

Insertion is classically used in particle-based state-space inference \citep[e.g., bootstrap particle filters:][]{GordonSalmondSmith1993, Kitagawa1996} for online latent state tracking but without retrospective refresh.
Iterated batch importance sampling \citep{Chopin2002} targets static parameters.
More recently, \cite{FongHolmes2021} applied mesh-based insertion for conformal uncertainty quantification in exchangeable Bayesian models, using add-one-in perturbation $g(\bm{\Theta}) = p(\bm{y}_{N+1} = \bm{y}^* \cond \bm{\Theta}, \bm{y}_{1:N})$ and target as conformity score via posterior predictive $p(\bm{y}_i \cond \bm{y}_{1:(N+1)})$.
This is approximated by weighted pre-perturbed posterior draws $\bm{\Theta}^{(r)} \sim p(\cdot \cond \bm{y}_{1:N})$ with $W^{(r)} \propto g(\bm{\Theta}^{(r)})$.
Although extreme $\bm{y}^*$ may be avoided under a large miscoverage level $\alpha$,
our framework can allow this, such as with sequential mesh-based approximation $g(\bm{\Theta};u) = p(\bm{y}_{N+1} = \bm{y}^*(u) \cond \bm{\Theta}, \bm{y}_{1:N})$ by rejuvenating when diagnosed as necessary.

\paragraph{Data deletion.}

Although not the focus of our application, deletion also arises naturally in approximate leave-one-out cross-validation and variants \citep[e.g.,][]{VehtariGelmanGabry2017}.
These ultimately remove data units by essentially modulating informativity $\gamma_i$.
Typically, intermediate distributions are treated as computational scaffolding but not of direct inferential interest \citep{HanGelman2026}.
{In our application, we are also interested in these intermediates, as the variation in configurations is informed by the user's qualitative query itself.}

\paragraph{Scenario analysis.}

\cite{BornnDoucetGottaro2010} examine the often-called (Bayesian) Lasso path \citep{ParkCasella2008}, which can be viewed via perturbation functions of the form $g(\bm{\Theta};u) = p_{\phi(u)}(\bm{\theta}_k) \cdot p_{\phi_k}(\bm{\theta}_k)^{-1}$ in Example \ref{example:prior};
the target and the modulated quantity is the regression coefficient, and the path is traced over different scale hyperparameters.

The meta-modeling lens would be relevant for scenario analysis outside of our specific application,
such as in Bayesian macroeconometric forecasting.
Conditional forecasts are often framed around counterfactuals \citep[e.g., what would happen to the system if some variable follows a specified path?:][]{WaggonerZha1999}.
\cite{McCrackenOwyangSekhposyan2021} present a setting in which a low-frequency outcome is forecast conditional on a hypothetical path for a higher-frequency input;
as new high-frequency data arrive, the forecast is routinely revised, which mirrors the sequential update we perform.

\paragraph{Sensitivity.}

From a procedural point of view, local sensitivity methods \citep[e.g.,][]{GiordanoBroderickJordan2018, Giordano2018, KallioinenPaanenBurknerVehtari2023} examine how posterior expectations shift under infinitesimal perturbations to hyperparameters $(\phi_k)_k$ or power coefficients.
These can be viewed as special cases of our perturbation framework, with $g(\bm{\Theta}; u) = \prod_k p_{\phi_k}(\bm{\theta}_k)^{\rho_k(u) - 1}$ encoding a one-step prior replacement,
and the target function is a chosen sensitivity metric, such as cumulative Jensen--Shannon divergence or empirical covariance \citep[Appendix B]{GiordanoBroderickJordan2018}.
The approach has already seen use in specialized domains; for example, in tipping‑point analyses in medical research, where sensitivity to priors is of interest \citep[e.g.,][]{OhigashiSugasawa2025}.

\section{Backtesting the forecasting model} \label{sec:4}
\subsection{{Election-level polling bias}} \label{sec:backtest:1}

To assess how the model responds to an industry-wide polling failure, we design a perturbation to the systemic polling-error prior.
The setting is the 2016 election, a year that remains a widely discussed example of polling error.
We evaluate the model 14 days before Election Day: a point at which most polls have already been aggregated, yet still early enough to remain meaningfully pre-election.
Specifically, we operationalize the total survey error \citep{Biemer2010} by simulating a uniform election-level polling error by shifting the prior mean of the systemic polling bias from zero to $\bm{b}(u_\ell) = b_0 u_\ell \bm{1}_S$, where $b_0$ corresponds to 0.75 times the prior standard deviation on the logit scale.
(As an illustration, on the probability scale from 0.5, this corresponds to shifting the average polling bias to roughly 0.51.)
\begin{align*}
    g(\bm{\Theta}; u_\ell)
        &= \text{MVN}(e_1, \ldots, e_S \cond \bm{b}(u_\ell), \bm{\Sigma}^{(\text{bias})})
        \cdot
        \text{MVN}(e_1, \ldots, e_S \cond \bm{0}, \bm{\Sigma}^{(\text{bias})})^{-1}
        ,
\end{align*}

\begin{figure}[h]
    \centering
    \includegraphics[width=\linewidth]{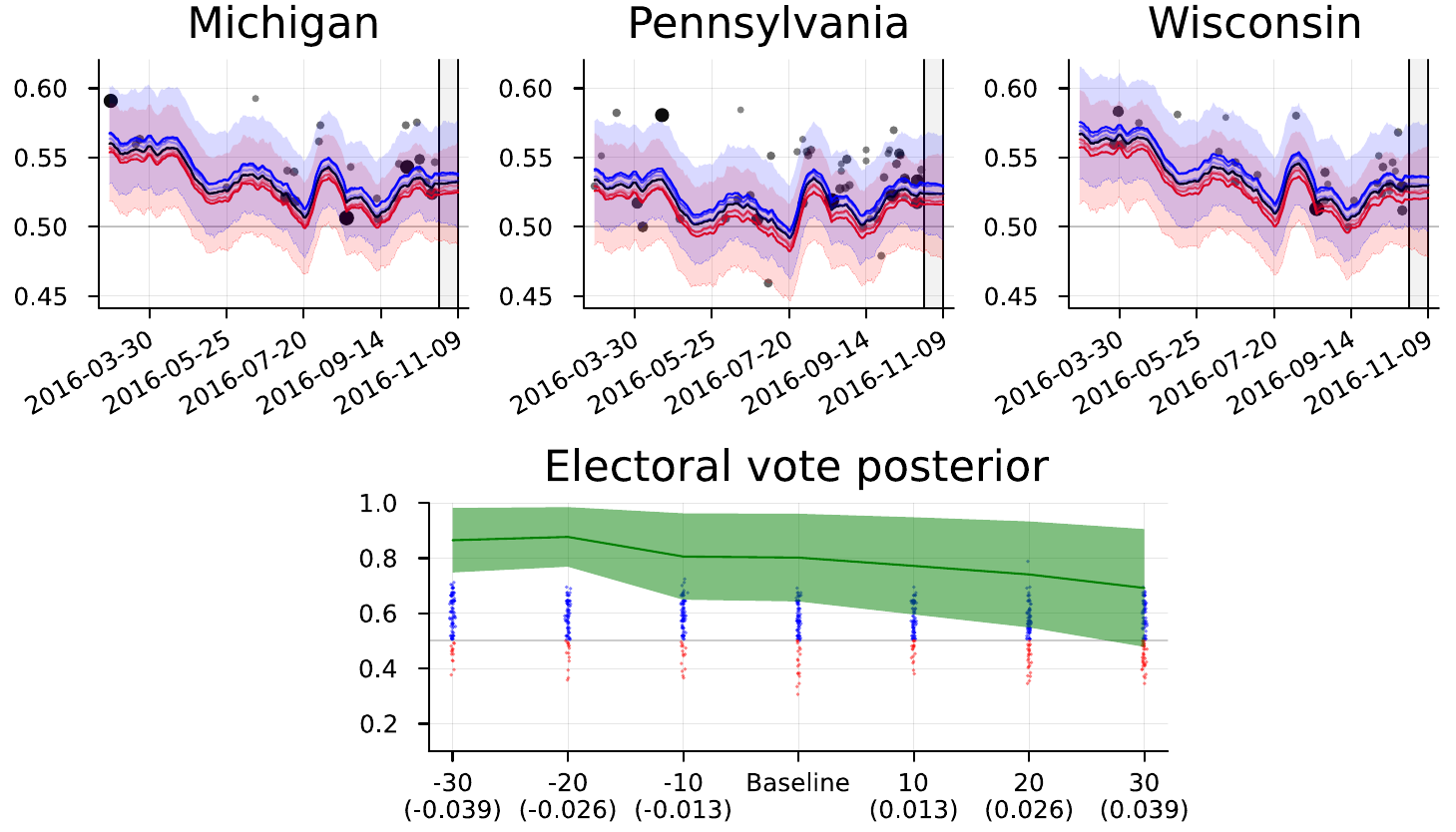}
    \vspace{-.2in}
    \spacingset{1.0} 
    \caption{\em
        Results from perturbation scheme \#1 (nonzero election-level polling bias: \ref{sec:backtest:1}).
        \textbf{Top}: state-level latent trend posterior means for Democratic Party support, before (\textbf{black line}) and after perturbation (\textbf{\textcolor{blue}{Democratic‑favoring}} and \textbf{\textcolor{red}{Republican-favoring}} polling bias direction) for Michigan, Pennsylvania, and Wisconsin.
        \textbf{Shaded region}: 90\% posterior credible interval.
        \textbf{Black dots}: individual poll results from March to 14 days before Election Day 2016.
        Dot size is proportional to polling sample size.
        \textbf{Bottom}: realized posterior of Electoral vote share.
        \textbf{\textcolor{Green}{Green line/band}}: posterior mean win probability and point-estimated unbiased variance.
    }
    \label{fig:polling-bias}
\end{figure}

Figure \ref{fig:polling-bias} first shows the change in forecasted Democratic win probability, namely in the Rust Belt swing states (i.e., Michigan, Pennsylvania, and Wisconsin).
As expected, imposing a systemic Democratic‑favoring bias \textit{a priori} produces a uniform downward shift.

The perturbation represents a correlated shock;
we also plot in Figure \ref{fig:polling-bias} the binary Electoral outcome as an aggregate measure.
The effects register directly, in the sense that even moderate shifts lead to substantial electoral votes being effectively relocated (see Figure \ref{fig:decomp} panel (b) also).
This demonstrates how the model's headline certainty (e.g., \dquote{80\%+ chance of Democratic victory}) can be highly sensitive to the assumption that systemic polling error is centered at zero:
which is an assumption the current model cannot identify or adjust for from the aggregate polling data alone (Section \ref{sec:model:q:bias}) and one that historical evidence shows is frequently violated \citep{Shirani-MehrRothschildGoelGelman2018}.

\begin{figure}[h]
    \centering
    \includegraphics[width=\linewidth]{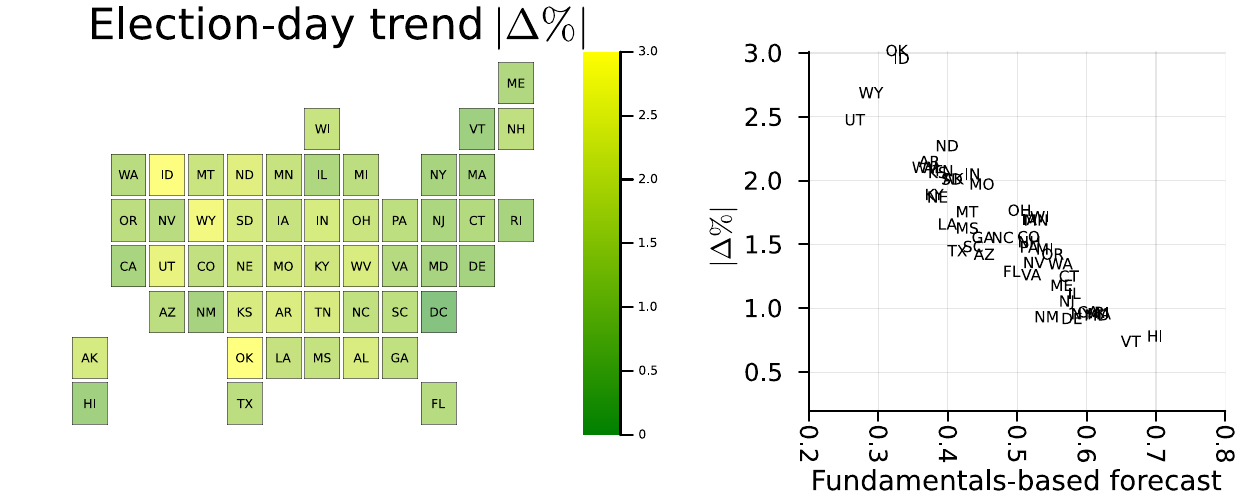}
    \vspace{-.2in}
    \spacingset{1.0} 
    \caption{\em
        Map of the (absolute) relative change in Election Day forecasts $\operatorname{logit}^{-1}(\bm{\mu}_{T})$.
        \textbf{Left}: heatmap of state‑level changes.
        \textbf{Right}: scatter plot against fundamentals-based state forecasts.
    }
    \label{fig:map}
\end{figure}

Figure \ref{fig:map} shows how the model responds over the map.
The latent trends do not move uniformly but instead exhibit asymmetric elasticity;
the model induces a shift that moves all states in the same direction, yet leaves states with fundamentals in favor of the Democratic party relatively insulated and vice versa.
This suggests partisan asymmetry with respect to the fundamentals in how the model absorbs correlated polling misses.
Uneven movement between states is expected, as states differ in sample size, polling volume, and information quality.
However, if the polls are contaminated by a uniform bias but the model assumes otherwise, then heterogeneous information (such as fundamentals here) can be used by the model to reallocate adjustments.
This is undesirable because the model's response to a correlated polling failure is driven by the structure of the prior, in that the model has no information about the underlying measurement error and must reconcile the conflict through other available channels.

\subsection{{Prior-data conflict and covariance}} \label{sec:backtest:2}

The previous exercise gives a natural segue into the next perturbation exercise, which examines the conflict more directly (see Section \ref{sec:model:q:corr}).
To see how the covariance weight influences downstream posterior inferences, we perturb upward and downward the covariance weight with $(\lambda \pm 0.24) \in [0.51, 0.99]$ about the default $\lambda = 0.75$.

\begin{figure}[h]
    \centering
    \includegraphics[width=\linewidth]{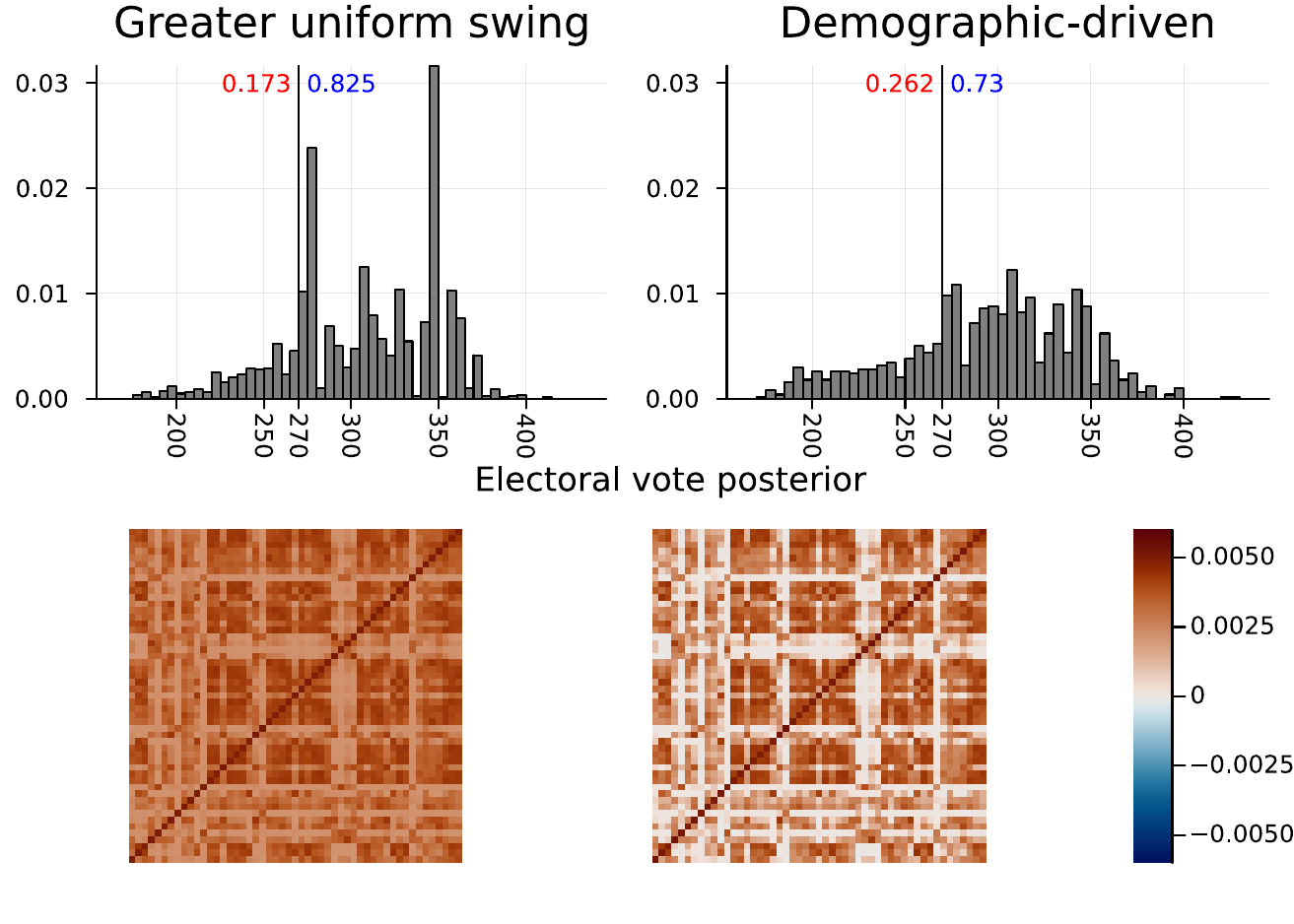}
    \vspace{-.2in}
    \spacingset{1.0} 
    \caption{\em
        Results from perturbation scheme \#2 (\ref{sec:backtest:2}):
        simulated Electoral vote posterior (Democratic vote share) under different correlation weights.
        \textbf{Left}: when $\lambda=0.75-0.24=0.51$.
        \textbf{Right}: when $\lambda=0.75+0.24=0.99$.
        \textbf{Bottom}: heatmap of the resulting perturbed state-by-state covariance matrix.
    }
    \label{fig:corr}
\end{figure}

Figure \ref{fig:corr} visualizes the scheme.
When greater weight is placed on the demographic-based correlation, states are encouraged to move in partially independent demographic clusters rather than as a single block, which leads to smoother and more diverse configurations of Electoral totals with fatter tails.
With greater equi-correlation, the movement is more uniform, and the distribution becomes much more concentrated.

Recall that the linear predictor on the logit scale is composed of six additive components (Eq.~\ref{model:eq:errors}):
the latent state trend, state-level polling bias, house effects, mode effects, population-screen effects, and the nonresponse adjuster.
Because the linear predictor is additive, perturbations redistribute contributions across components, but as we perturb the covariance-weight, we can track how the model re-distributes across these for each state by averaging across the state-polls.

\begin{figure}[h]
    \centering
    \subfloat[Reduced equi-correlation \eqref{sec:backtest:2}]{\includegraphics[width=.46\linewidth]{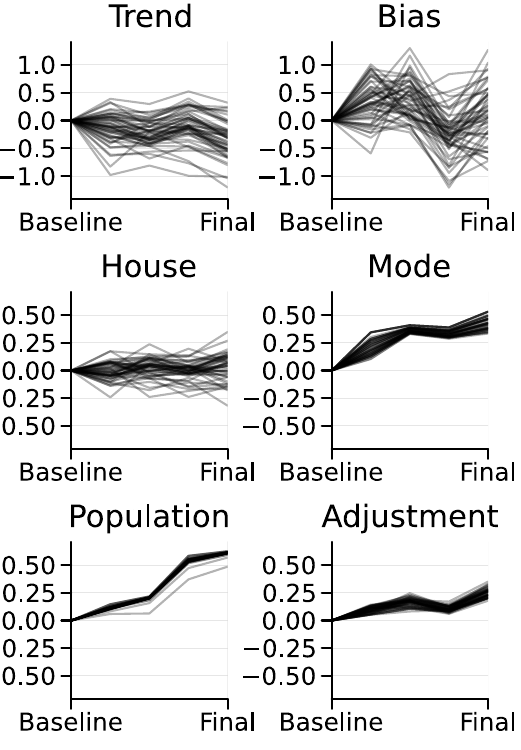} \label{fig:decomp:2}} \qquad
    \subfloat[Greater Democratic-favoring bias \eqref{sec:backtest:1}]{\includegraphics[width=.46\linewidth]{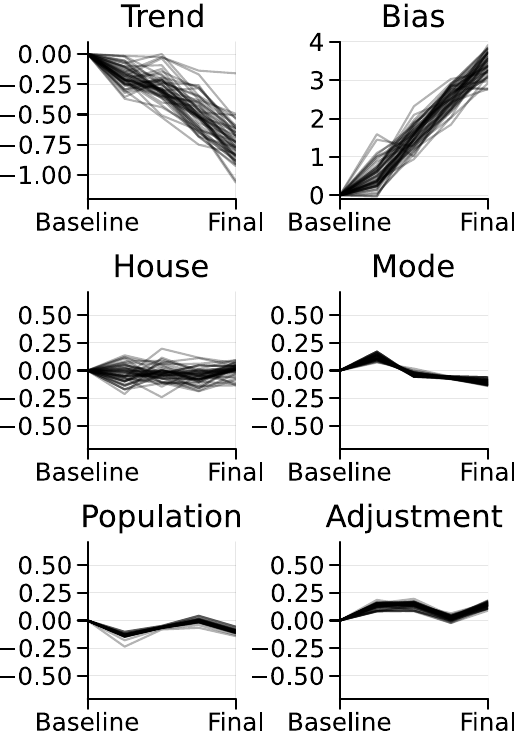} \label{fig:decomp:1}}
    \spacingset{1.0}
    \caption{\em
        Results from perturbation scheme \#1 (panel \ref{fig:decomp:1}) and \#2 (panel \ref{fig:decomp:2}):
        state-by-state decomposition of the logit-scale linear predictor.
        Each line shows, for each state, six additive components averaged across available state-polls to show how the model re-allocates:
        latent trend at $T$ (\dquote{Trend}), state‑level polling bias (\dquote{Bias}), house effects (\dquote{House}), mode effects (\dquote{Mode}), population-screen effects (\dquote{Population}), and the non‑response adjuster (\dquote{Adjustment}).
        \textbf{Horizontal axis}: step of perturbation.
        \textbf{Vertical axis}: posterior mean(s) on the logit scale, centered about the baseline value.
    }
    \label{fig:decomp}
\end{figure}

The decomposition results are displayed in Figure \ref{fig:decomp} panel (a).
For reference, results from the previous polling bias perturbation are shown in panel (b).
(b) Under a uniform positive polling bias, the model primarily routes the discrepancy into the polling bias term (understandably).
Others, except for the latent state trajectory, do not vary discernibly.
(a) When the equi-correlation contribution is reduced, the model allocates variation to the mode and population intercepts, visible by how its posterior mean increases.
The opposite perturbation gives symmetric effects.
When the spatial covariance is prevented from absorbing the prior-data conflict, the model is forced to redirect the unexplained variation into the remaining non-spatial but nationwide channels.

This illustrates two related risks.
First, a seemingly innocuous choice of covariance weight can unintentionally limit the model's ability to learn from the data;
a higher subjective setting could lead the polling miss to be absorbed through other terms that were never intended to explain such variation.
Relatedly, the sensitivity of these methodological terms to the covariance weight suggests that the model can mis-infer genuine variations from polling metadata information, as the intercepts no longer represent what they were designed to capture, and is crucially intended to ultimately be discarded when generating vote-share predictions (see Section \ref{sec:2:1}).
The misattribution, however, is reasonable given the model's current structure;
this in turn suggests that the model must be supplied with information beyond aggregated polling data to mitigate these risks of structural misattribution.

\subsection{{Quantifying the asymmetries in poll assimilation}} \label{sec:backtest:3}

The results thus far have highlighted how the model inevitably misexplains variation such as via the (partly-uniform) covariance structure or other global methodological intercepts as second-best channels for absorbing discrepancies (Section \ref{sec:backtest:2}).
One manifestation of this is the unintended partisan asymmetry observed in the posterior (Section \ref{sec:backtest:1}).
Here, we shall concretely quantify further asymmetries through a concrete scenario analysis.

We further study the model's geographic covariance by examining how a structured pattern of injected polls alters the forecast and propagates through.
Specifically, we inject identical \textit{adversarial} polls into each state $s$, of sufficiently large sample size $n_i=800$.
The poll is \textit{adversarial} in the sense that its implied support rate $y_i/n_i$ is set to $\operatorname{logit}^{-1}(\mu_{T,s}) + 0.05$ (or $\operatorname{logit}^{-1}(\mu_{T,s}) - 0.05$) when the current Election Day forecast $\operatorname{logit}^{-1}(\mu_{T,s})$ is below (or above) 0.5, so that the injected poll meaningfully deviates from the model's current trajectory.
The poll is gradually introduced into the posterior by holding $y_i/n_i$ fixed while increasing/rounding both $y_i$ and $n_i$, thus preserving binomial support.
The setup here is 60 days out (rather than the earlier 14-days just‑before horizon), where the model has meaningful room to adjust its trends.
Repeating this across all 50 states allows us to directly quantify how localized shocks are absorbed and propagated through the model.

\begin{figure}[h]
    \centering
    \includegraphics[width=\linewidth]{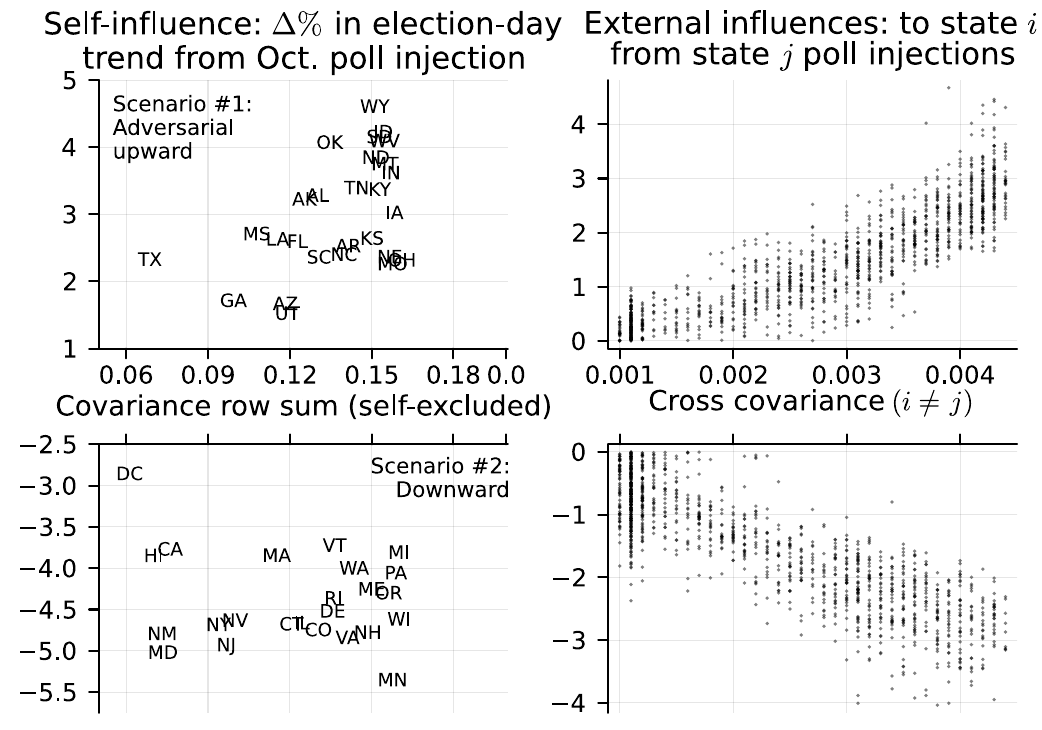}
    \vspace{-.2in}
    \spacingset{1.0} 
    \caption{\em
        Results from perturbation scheme \#3 (hypothetical adversarial poll injection: \ref{sec:backtest:3}).
        \textbf{Left}: state-by-state relative change in Election Day forecasts, plotted against each state's pre-computed centrality (i.e., sum of its nonnegative covariances with all other states).
        \textbf{Right}: cross-state external influence, shown as the relative change in Election Day forecasts for other states versus their cross‑covariance values.
        \textbf{Note}: all covariance entries are nonnegative by construction (see Section \ref{sec:model:q:corr}).
    }
    \label{fig:transmission}
\end{figure}

Figure \ref{fig:transmission} plots each state's latent absorption, against demographic centrality (i.e., sum of its nonnegative covariances with all other states).
We show two series: one for a $+5$-point injected poll and one for a $-5$-point poll.
When the injected poll is above (below) the state's current trend, more central states show larger positive (negative) shifts.
In addition, states with stronger cross-covariance exert more external influence (i.e., absorptions in other states).
Taken together, the covariance acts more as a transmission mechanism than a stabilizing prior.

As such, a single noisy poll in a high‑centrality state can propagate broadly and generate broad shifts across states that reflect the geometry of the prior rather than that of the data.
This is undesirable due to the fact that the sampling mechanism for polls is non‑ignorable (i.e., poll availability and composition carry information about the underlying process: Section \ref{sec:model:q:asymmetry}), yet the model as it stands is deficient in such information (Section \ref{sec:model:q:bias}).
Moreover, even modest mis-tuning of the covariance, as possibly a result of an innocuous data-wrangling choice (see Sections \ref{sec:model:q:corr} and \ref{sec:backtest:2}), can exacerbate such downstream overreactions.

\subsection{{Implications}} \label{sec:backtest:implications}

Unmodeled nonsampling errors can introduce prior-data conflict, and the model resolves it through unintended adjustment pathways that are dictated by subjective structural assumptions embedded in the prior and covariance design.
The model, as currently specified, does not have a sufficient structure to represent these sources of error.
For example, the absence of a poll in a given state-day cell is not treated as informative missingness that should be jointly modeled and marginalized over.

Immediate remedies that do not require dramatic redesign would at minimum incorporate information about such absence, such as nonzero polling bias, and relax the rigidity of the covariance structure.
Without such a structure, the model inevitably misfits the data, which it then reconciles through pathways that are highly sensitive to the assumed structure, such as covariance and other global terms.

On the other hand, although such minor modifications to the model architecture can help, they are likely insufficient on their own.
Much more substantial is augmenting the workflow with additional sources of variation, whether directly into the model or at a broader stage of the forecasting process.
For example, forecasters should incorporate additional sources of information beyond aggregated polling alone within the polling model, and not treat each election as an independent draw, by incorporating time-varying information across election cycles.
It would further be sensible to incorporate demographic information through sensible frameworks such as multilevel regression and poststratification \citep[a.k.a., MRP:][]{GelmanLittle1997, GhitzaGelman2013}, such as with post-stratification tables for each geographic unit that explicitly partition the electorate into demographic strata, rather than relying on heuristic covariance‑based adjustments.
These illustrate the broader point that the sequential sampling scheme allowed us to see these issues concretely and motivate refinements to the model.

\section{Discussion} \label{sec:5}
Real-time Bayesian dynamic election forecasting presents a unique methodological and practical challenge at the intersection of daily posterior updates, time-specific scenario and sensitivity analysis, and collaborative communication.
The core issue is that the model is never truly complete;
rather, it is a living object, revised routinely as new data and events arrive.
To move toward a model that is reliable in real-time, we sought a framework that accommodates timely updates and qualitative diagnostic queries (e.g., Q1--Q3).

We developed a meta-modeling framework, paired with a sequential sampling scheme, to enable efficient model and forecast revision in real-time forecasting environments where data and assumptions evolve continuously.
The former meta-modeling component generalizes the original forecasting model by explicitly operating over fixed model quantities (e.g., hyperparameters) to enable interpretable scenario and sensitivity analyses.
The latter sequential sampling scheme addresses the computational inefficiencies of brute force MCMC by taking advantage of the sequential proximity of posterior distributions along a perturbation path.

Our case study revealed several undesirable features of the model and crucially motivated improvements to the model, as detailed in Section \ref{sec:backtest:implications}.
Moreover, we show in Section \ref{sec:appendix:sim} that the sequential sampling scheme enables such analyses at a fraction of the computational cost compared to brute force MCMC, while yielding posterior approximations that are nearly indistinguishable.

We conclude by pointing to several directions for future work.
Although our case study focused on real-time Bayesian election forecasting, the framework is readily extensible to other domains involving sequential revision.
For example, macroeconomic monitoring often features asynchronous data arrivals and frequent updates to select indicators \citep[e.g.][]{McCrackenOwyangSekhposyan2021}, which makes the data-value perturbation and insertion strategies directly relevant.
More broadly, in general statistical analyses, data-value perturbation may be used to encode aspects of human decision-making in the {\em analysis pipelines} (e.g., preprocessing choices, data-cleaning decisions, or model-selection heuristics), to gauge the stability of (posterior) inferences \citep{RewolinskiYu2025}.
On the methodological side, the approach in Section \ref{sec:3:2} could be useful for exploring broader model changes in Bayesian settings: e.g., inclusion of interaction terms in hierarchical regression.
Such step‑by‑step exploration of model structure, combined with fast sequential simulation-based inference, would support flexible, efficient, and interpretable Bayesian workflows.

\if0\blind
{

\section*{Acknowledgments}
We thank Ben Goodrich and Dan Rosenheck for collaboration.
GH was supported by JST SPRING Grant Number JPMJSP2108.
AG was partially supported by the Office of Naval Research.


\section*{Disclosure Statement}
There are no competing interests to declare.

\section*{Data Availability Statement}
The data and accompanying code are available at \href{https://github.com/geonhee619/SMC-Sense}{https://github.com/geonhee619/SMC-Sense}. These data were derived from \href{https://github.com/TheEconomist/us-potus-model}{https://github.com/TheEconomist/us-potus-model}.

} \fi

\if1\blind
{
} \fi

\clearpage

\appendix \section{Supplementary} \label{sec:appendix}

\subsection{{Application-oriented simulated validation}} \label{sec:appendix:sim}

This subsection presents a simulated evaluation of the accuracy and computational efficiency.
There is no shortage of choices for perturbations;
we focus specifically on the perturbation schemes adopted in the main text.
\begin{enumerate}[(\#1)]
    \item Section \ref{sec:backtest:1}: modulating the election-level state-by-state correlated polling error $(e_s)_s$ location parameter, which by default is zero-centered.
    
    \item Section \ref{sec:backtest:2}: modulating the covariance weight $\lambda$ in the fixed state-by-state covariance matrix $\lambda\bm{R}+(1-\lambda)\bm{R}_1$ which is used in the workflow to construct $\bm{\Sigma}^{(\mu)}$ and $\bm{\Sigma}$.
    
    \item Section \ref{sec:backtest:3}: injecting a fictitious state-level poll over all states.
    For illustration, we focus on the case when $(s[i],h[i])=(\texttt{Ohio}, \texttt{ABC})$.
    
    \item Same as above, but when $h[i]=\texttt{Pew}$.
    Fixing $(s,h)$ as such is natural and fair since the analyses are parallelizable for a prescribed metadata, both for SMC and bruteforce MCMC.
\end{enumerate}

Across all scenarios, we evaluated the accuracy of the approach by comparing to MCMC draws as a reference, each obtained independently from the corresponding perturbed posterior (i.e., bruteforce).
Runtime is also recorded for comparison.
For each case, SMC is run with 30 discrete mesh points based on the absence of warning outputs (see Section \ref{sec:3:3:diagnostics}).
Moderate parallelization (4 threads) is used during particle rejuvenation (see Section \ref{sec:3:3:parallel}).
The MCMC kernel (chosen as Hamiltonian Monte Carlo) is applied 3 times, informed by baseline MCMC draw autocorrelation following \cite{HanGelman2026}.
For the reference MCMC, we run 1200 iterations with a burn-in of 200.
The 1000 posterior draws given $\bm{\delta}_t(0)$ at time $t$ (in \ref{sec:3:2}) serve as input to SMC for all subsequent analyses.
The ESS threshold is $1/2$, which is typical.

\begin{figure}
    \centering
    \includegraphics[width=\linewidth]{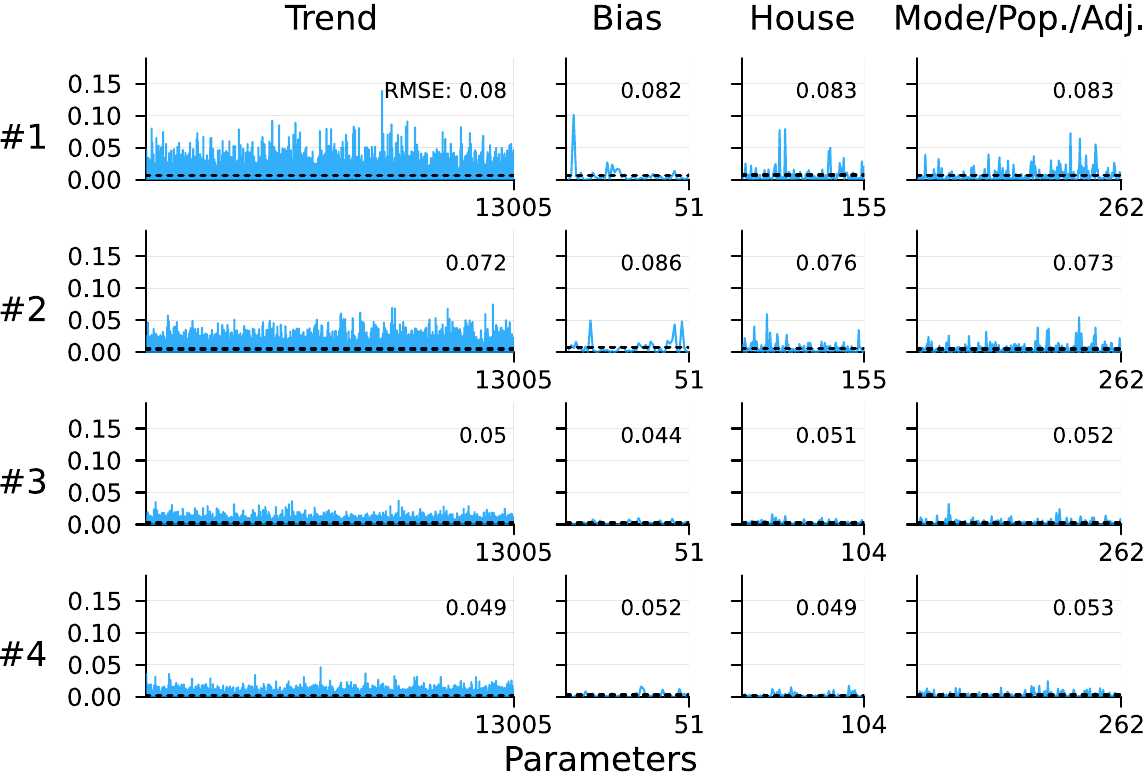}
    \vspace{-.3in}
    \spacingset{1.0} 
    \caption{\em
        Comparison between brute force MCMC and sequential sampling (SMC).
        \textbf{\textcolor{blue}{Blue line}}: vector of squared errors of the posterior mean estimates in the final perturbation step.
        \textbf{Dotted line} and \dquote{\textbf{\textit{RMSE}}}: root mean squared error over the vector.
        \textbf{Note}: error is scaled by the posterior standard deviation obtained from MCMC.
        From top to bottom, perturbations are:
            (\#1) modulating the zero-centered polling error $(e_s)_s$ location (\ref{sec:backtest:1});
            (\#2) modulating the covariance weight $\lambda$ in the state covariance upward (\ref{sec:backtest:2});
            (\#3) injecting a fictitious adversarial poll with $(s[i],h[i])=(\texttt{Ohio}, \texttt{ABC})$ (\ref{sec:backtest:3});
            (\#4) with $(s[i],h[i])=(\texttt{Ohio}, \texttt{Pew})$ (also \ref{sec:backtest:3}).
    }
    \label{fig:sim:accuracy}
\end{figure}

{Figure \ref{fig:sim:accuracy} presents the squared absolute error between the posterior mean estimates for the parameter vector in the final perturbation step (i.e., the most distant posterior).
The error is scaled by the posterior standard deviation obtained from the MCMC draws.}
The number of parameters is 14693 at 14 days out (\#1 and \#2 in the Figure), and 14008 at 60 days out (\#3 and \#4 in the Figure).
All parameters are in the noncentered and unconstrained parameterization of standard normal priors.
The Figure shows that the vast majority of the parameters have absolute squared error at most 0.05 of the posterior standard deviation, which indicates that there are only practically negligible deviations between the two approaches in their respective perturbation schemes.

\begin{figure}
    \centering
    \includegraphics[width=\linewidth]{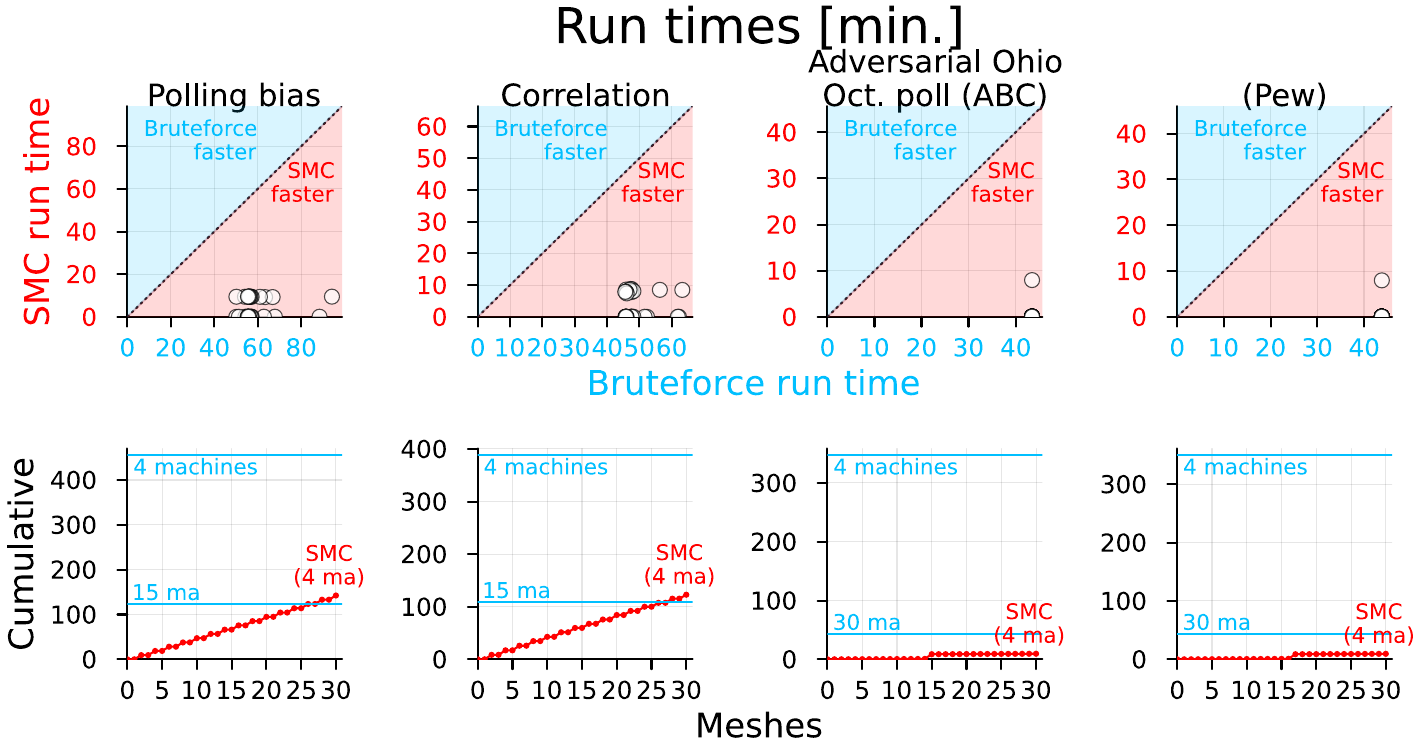}
    \vspace{-.2in}
    \spacingset{1.0} \caption{\em
        Run time comparison between brute force MCMC and sequential sampling (SMC).
        \textbf{Top}: Comparison per perturbation step.
        \textbf{Bottom}: Cumulative run time across all 30 steps.
        See Figure \ref{fig:sim:accuracy} caption for succinct descriptions.
        Moderate parallelization (4 threads) is used.
        The maximum number of parallel chains required for brute force MCMC to outperform the sequential approach is also shown.
    }
    \label{fig:sim:runtime}
\end{figure}

Figure \ref{fig:sim:runtime} compares the run time per perturbation step (i.e., per mesh point) and cumulative (i.e., the total time to approximate the full modulation scheme).
As for the former, sequential sampling is consistently faster, especially in cases where particle rejuvenation is not triggered, as indicated by the concentration of points near the horizontal axis;
reweighting via density/likelihood evaluation is inexpensive relative to generating a new set of draws based on multiple gradient computations.
The cumulative run time is also lower under equal and moderate parallelization (4 threads in our setup).
For reference, we also report the hypothetical minimum number of parallel machines required (up to the mesh count of 30) using brute force MCMC to match those of sequential sampling.
Depending on the perturbation scheme, achieving comparable performance would require much greater computational resources.

\newpage

\subsection{Practical implementation notes}

\subsubsection{An example} \label{sec:3:3:example}

Our implementation is carried out in Julia.
In brief, the workflow proceeds as follows.
\begin{enumerate}[(a)]
    \item Specify the Bayesian model using \texttt{Stan} \citep{CarpenterGelmanHoffmanLeeGoodrichBetancourtBrubakerGuoLiRiddell2017}, as in the forecasting workflows of \cite{HeidemannsGelmanMorris2020} and \cite{GelmanGoodrichHan2024}; see also \cite{morris2020}.
    
    \item Define a sequence of \texttt{Stan} meta-models.
    
    \item Traverse the meta-models sequentially by accessing the model's log density and its gradient via \texttt{BridgeStan} \citep{RoualdesWardCarpenterSeyboldtAxen2023}.
\end{enumerate}
The workflow is modular.
In (a), the user first obtains the baseline draw.
In (b), for example, varying a hyperparameter (\ref{example:prior}) involves modifying the \texttt{data} block.
The following is an illustrative code.
\begin{lstlisting}[language=Python, caption={}]
model_0 = StanModel("model.stan", "data.json")  # baseline model
data_0  = load_json("data.json")                # baseline configuration

for (i, scalar) in enumerate([1.1, 1.2, 1.3])
    data_1 = data_0 |> deepcopy       # inherit original configurations
    data_1["mu_b_T_scale"] *= scalar  # new configuration
    # more can be specified by user
    
    save_as_json(data_1, "perturbed_data_$i.json")               # save
    model_1 = StanModel("model.stan", "perturbed_data_$i.json")  # target
end
\end{lstlisting}
In (c), the ability to evaluate log-density and its gradient satisfies the requirements of SMC without requiring code rewrites for each perturbation scheme.
Sequential sampling is then performed as follows.
\begin{figure}
\centering
\begin{lstlisting}[language=Python, caption={}]
SMCS_Stan(
    draws_0,                           # baseline draw
    models_vec |> return_LogDensities  #  sequence of meta-models
);
\end{lstlisting}
\includegraphics[width=.7\linewidth]{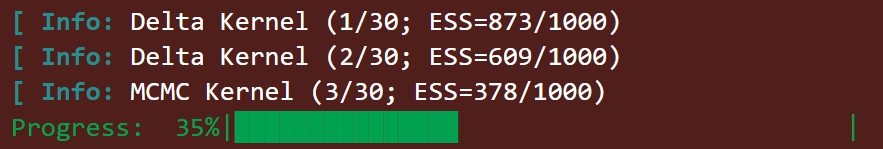}
\spacingset{1.0} \caption{\em
    Illustrative intermediate output.
    Given a sequence of meta-models and access to log density, Markov kernel selection and diagnostics (e.g., ESS and $\hat{k}$) are automatically performed as described in Section \ref{sec:3:2} to ensure sufficient sample diversity.
    In the above, rejuvenation was triggered due to $ESS < 500$ (i.e., below the 0.5-of-$R$ threshold) at step \#3 of 30.
} \label{fig:output-sample}
\end{figure}

\subsubsection{Parallelization} \label{sec:3:3:parallel}

On any given day, distinct and branching perturbative sequential updates are inherently parallelizable,
provided sufficient distributed computational resources.
Parallelization also applies within the perturbation paths.

When triggered, the MCMC kernel is applied independently per particle for rejuvenation (i.e., short runs of many chains).
In principle, assuming adequate resources, the rejuvenating computation can be fully distributed across particles, so that we can reduce the effective cost to that of a single-short-chain MCMC up to communication.
We actively leverage the parallelizability in our implementation (and recommend doing so in other applicable contexts similarly).

\subsubsection{Failure diagnostics} \label{sec:3:3:diagnostics}

It is critical to diagnose when a given approximation may be unreliable, so that computational effort is not wasted on perturbations that will ultimately be discarded.
For example, consider the illustrative but deliberately extreme one-step scheme
$$h(\bm{\Theta};u_1) = \mbox{binomial}(y^* = 89,\!999 \cond n = 90,\!000, \pi_i),$$
which corresponds to a hypothetical poll with sample size $n = 90,\!000$ and two-party support $y^* = 89,\!999$.
Such influential data would induce substantial posterior shifts, and any finite weighted set of baseline draws would likely fail to adequately approximate the updated posterior.

A simple remedy is then to incrementally approach the target (e.g., with a power scaling coefficient for gradual data injection),
along with the safe strategy to preventively prepare a sufficiently fine mesh,
since Algorithm \ref{alg:smc} avoids the costly MCMC kernels wherever possible.
However, determining whether the mesh is fine enough may not be obvious {\em a priori}.
Given the incremental structure of importance weights across mesh transitions, we implement the generalized Pareto shape diagnostic \citep{VehtariSimpsonGelmanYaoGabry2024} on the incremental component.
Diagnostic warnings indicated by $\hat{k} > 0.7$ would signal the need to prepare finer mesh points to ensure a reliable approximation.

\subsubsection{{Choice of schemes specific to SMC}} \label{sec:3:3:smc}

We acknowledge that alternative schemes may be preferable in other contexts;
our algorithm is a specific instantiation of a more general framework.

\paragraph{Resampling schemes.}

A variety of resampling schemes may be used.
Our implementation adopts multinomial resampling, in which the particle indices are redrawn proportionally to $(W_\ell^{(1)}, \ldots, W_\ell^{(R)})$.
An alternative is residual resampling, which first allocates deterministic copies according to the integer parts of the weights and then samples the remainder stochastically \citep[sec.~4]{LiuChen1998}.
Stratified resampling \citep{Kitagawa1996} draws one sample per stratum of the unit interval to reduce variance.
Systematic resampling \citep{CarpenterCliffordFearnhead1999} uses a single random offset and evenly spaced points.
An overview is given in \cite{ChopinPapaspiliopoulos2020} Section 9.

\paragraph{Temperature plan.}

In our setting, the temperature schedule is predetermined, as this is part of an application-specific informed choice.
In other applications, however, one may prefer a tuning‑free approach, such as when only the final draw at the end of the temperature plan is of interest, in which case the intermediate temperatures can/should be selected automatically.
Such is known as adaptive SMC or tempering, in which the next distribution is chosen adaptively so that successive targets remain sufficiently close, according to some criterion.
We refer to \cite{JasraStephensDoucetTsagaris2011}, \cite{ZhouJohansenAston2016} and \cite{BeskosJasraKantasThiery2016} for detailed descriptions and applications.



\bibliographystyle{apalike}
\bibliography{ref.bib}

\end{document}